\documentclass[prd,onecolumn,showpacs,floatfix,superscriptaddress,nofootinbib]{revtex4}
\usepackage{graphicx}
\usepackage{epsfig}
\usepackage{bm}
\usepackage{amssymb}
\usepackage{float}
\usepackage{amsmath}
\usepackage{dcolumn}
\usepackage{cancel}
\usepackage[colorlinks]{hyperref}
\usepackage[usenames,dvipsnames]{color}
\hypersetup{
     breaklinks=true,
        pdfstartview={FitH},     colorlinks=true,
    linkcolor=blue,
     citecolor=red,
    filecolor=magenta,
    urlcolor=blue,
    anchorcolor=green,
    linktocpage=true
}

\def\doi{http://doi.org}

\def\r{\mathrm{r}}

\def\m{\mathrm{m}}

\def\m{ \mu}
\def\n{ \nu}

\def\r{ \rho}

\def\na{\nabla}

\def\be{\begin{equation*}}
\def\ee{\end{equation*}}

\def\M{\mathcal{M}}

\def\G{\mathcal{G}}
\def\bex{\bar{B}_{\text{ext}}}
\def\bb{\bar{B}}

\def\m{ \mu}
\def\n{ \nu}

\def\r{ \rho}

\def\m{ \mu}
\def\n{ \nu}

\def\r{ \rho}

\def\na{\nabla}
\def\p{\phi}

\def\hg{\hat{g}}
\def\be{\begin{equation}}
\def\ee{\end{equation}}
\def\bea{\begin{eqnarray}}
\def\eea{\end{eqnarray}}
\def\sdg{\sqrt{-g}}
\def\sdgb{\sqrt{-\hg}}
\def\tm{T_{\text{min}}}
\def\tmv{{T_{\text{min}}^0}}

\def\leq{\leqslant}
\def\tc{\tau_{c}}
\def\tbc{\bar{\tau}_{c}}
\def\ts{T_{\star}}

\begin{document}

\title{Hairy Black Holes in Disformal Scalar-Tensor Gravity Theories}

\author{Cristian Erices}
\email{cristian.erices@ucentral.cl}
\affiliation{Universidad Central de Chile, Vicerrectoría Académica, Toesca 1783, Santiago, Chile.}
\affiliation{Universidad Católica del Maule, Av. San Miguel 3605, Talca, Chile.}

\author{Pantelis Filis}\email{pantfx31@gmail.com}
\affiliation{Physics Division,
National Technical University of Athens, 15780 Zografou Campus,
Athens, Greece.}
\author{Eleftherios Papantonopoulos}
\email{lpapa@central.ntua.gr} \affiliation{Physics Division,
National Technical University of Athens, 15780 Zografou Campus,
Athens, Greece.}


\begin{abstract}
We show that the no-hair theorem for scalar-tensor theories with bi-metric structure can be evaded. We find that hairy black hole solutions in the presence of an electric charge admit AdS, flat or dS asymptotics with spherical, flat, or hyperbolic base manifolds. Spherically symmetric, asymptotically flat black holes and asymptotically AdS configurations with any horizon topology are compatible with a regular scalar field on and outside the event horizon. The latter presents a rich thermodynamic behavior induced by the disformal factor that enters as a coupling parameter in the theory. In the grand canonical ensemble, there is an interplay of stability and first-order phase transitions between thermal AdS, the hairy black hole, and the Reissner-Nordstr\"om-AdS black hole, whose thermodynamic phase space resembles a solid-liquid-gas system, with an electric potential playing the role of pressure. In close analogy, there is a triple point where the three phases coexist, being equally probable.
\end{abstract}

\maketitle

\section{Introduction}

As it was discussed by Bekenstein in \cite{Bekenstein:1992pj} to describe gravitation, one may need two geometries.
One of these describes gravitation, while the other defines the geometry in which matter describes the gravitational dynamics.
This approach is necessary if one wants to formulate a modified theory of gravity. To avoid
 conflicts with the tests of general relativity (GR), one has to use a Riemannian metric
$g_{\mu \nu}$ describing  the geometry's dynamics and in order to see the effect from the
departure from standard GR one has to introduce a relation between $g_{\mu \nu}$ and the
physical geometry on which matter propagates.

Then he introduced the following Riemannian metric
\begin{equation}
ds^2= \hat{g}_{\mu \nu}  dx^{\mu} dx^{\nu} \equiv \Big{(} g_{\mu \nu} A +L^2 B \partial_{\mu} {\phi} \partial_{\nu}{\phi}\Big{)} dx^{\mu} dx^{\nu}~,
\label{geff}
\end{equation}
where $L$ is a length scale and $A$ and $B$ are in general
functions of the scalar field $\phi$. The physical metric $g_{\mu \nu} $ and the matter metric $\hat{g}_{\mu \nu} $ are related by a conformal and a disformal transformation. The physical understanding of the relation of the two metrics in \eqref{geff} was presented in \cite{Bekenstein:1992pj}. When $B=0$ a conformal transformation relates these two metrics. This transformation leaves all shapes invariant and stretches equally all spacetime directions. When $B\not= 0$ we have a disformal tranformation which its effect is that the stretch in the direction parallel to $\partial_{\mu} \phi$ is by a different factor from that in the other spacetime directions and shapes are distorted. To see the physical context which is introduced by the disformal transformation the field equations should be written at the outset with the metric $\hat{g}_{\mu \nu} $.

This bi-metric scalar-tensor gravity theory developed by Bekenstein was employed in cosmology to explain observational results and study the propagation of gravitational waves. In \cite{Clayton:1999zs,Clayton:2001rt} it was argued that the contribution from the scalar field in the metric ${\hat g}_{\mu\nu}$ can generate acceleration in the expansion of the universe, without negative pressure and with zero cosmological constant and gravitational waves will
propagate at a different speed from non-gravitational waves. Furthermore, they found that the gravitational waves and matter waves have different propagation speeds in this bi-metric structure. They studied a model in which ordinary matter is coupled
to the matter metric $\hat{g}_{\mu \nu} $ of \eqref{geff} with $A=L^2=1$. All matter fields except the scalar field $\phi$
propagate in the geometry described by $\hat{g}_{\mu \nu} $ while matter and radiation will propagate along geodesics determined by this geometry and obey the equivalence principle. A result of this consideration is that the speed of gravitational wave propagation is found to be significantly different from the speed of matter waves and photon propagation in the early universe. In \cite{Magueijo:2000au} stars and black holes in varying speed of light theories were studied.

Disformal transformations were used in the study of various scalar-tensor gravity theories. One of them is the most general scalar-tensor theory leading to second-order field equations in four dimensions is the Horndeski theory. In \cite{Bettoni:2013diz} it was shown that disformal transformations play, for the Horndeski theory, a similar role to conformal transformations for scalar-tensor theories. Disformal transformations were used in higher-order scalar-tensor Horndeski theories to study the stability of these theories, and the absence of ghosts \cite{Achour:2016rkg,Deffayet:2020ypa}. In these theories, cosmological perturbations were studied in \cite{Tsujikawa:2014uza}, and it was shown that both curvature and tensor perturbations on the flat isotropic cosmological background are invariant under the disformal transformation.

The formation of compact objects in a class of scalar-tensor theories with disformal coupling to matter was studied in \cite{Minamitsuji:2016hkk}. A minimal model of a massless scalar-tensor theory was proposed, and an investigation was carried out of how the disformal coupling affects the spontaneous scalarization of slowly rotating compact objects. In \cite{BenAchour:2019fdf} starting from suitable seed solutions in degenerate higher-order scalar-tensor theories, new solutions were discussed using disformal field transformations, and also hairy black hole solutions were obtained.

The aim of this work is to find hairy black holes in bi-metric theories in which the two metrics are connected by a disformal transformation. In this study we will let in (\ref{geff}) $A=L^2=1$ and $B $ to be constant independent of the scalar field $\phi$. As we mentioned, when $B=0$ the two metrics are connected by a conformal transformation. Black hole solutions where the scalar field is conformally coupled to gravity were found by Bocharova, Bronnikov and Melnikov and independently by Bekenstein, called BBMB black hole \cite{BBMB}. The spacetime of this solution is the extremal Reissner-Nordstr\"om (RN) spacetime, but the scalar field diverges at the black hole horizon. Later, a cosmological constant was introduced in \cite{Martinez:2002ru} and also a quartic scalar potential that respects the conformal invariance and a simple relation between the scalar curvature and the cosmological constant was generated. In this case, the scalar field does not diverge at the horizon of the black hole. Black hole solutions in scalar-tensor theories were found
\cite{Kolyvaris:2009pc}-\cite{Charmousis:2014zaa}
 and even boson stars and black holes in bi-scalar extensions of Horndeski theories were studied in \cite{Brihaye:2016lin}. Recently,  black hole solutions with a conformally coupled scalar field  were studied in \cite{Karakasis:2021rpn,Cisterna:2021xxq,Erices:2017izj,Cisterna:2018hzf}. Black hole thermodynamics in AdS space has been also studied for hairy solutions in higher dimensions \cite{Arratia:2020hoy,Cisterna:2014nua}.

The primary motivation of this work is to study if we can evade the non-hair theorem in scalar-tensor theories with bi-metric structure in four dimensions, and if we can generate hairy black holes with a regular scalar field on the horizon and beyond. We will consider a scalar field $\phi$ coupled to the physical metric $g_{\mu\nu}$ and an electromagnetic field coupled to the matter metric $\hg_{\mu\nu}$. Solving the field equations we will show that the theory admits hairy black hole solutions with regular scalar field on and outside the horizon. Since the disformal factor $B$ is a coupling constant, it defines an effective cosmological constant and the spacetime can be asymptotically flat, dS or AdS. Studying the thermodynamics we will show that the solution has a rich thermodynamic behavior. In the grand canonical ensemble, there is an interplay of stability and first-order phase transitions between the hairy black hole, the Reissner-Nordstr\"om-AdS black hole and thermal AdS, whose thermodynamic phase space resembles a solid-liquid-gas system, with an electric potential playing the role of pressure. In close analogy, there is a triple point where the three phases coexist, being equally probable.

This work is organized as follows. In Section \ref{sex2} we present the general setup of the model. In Section \ref{sex3} we perform the thermodynamical analysis using the Euclidean approach. In Section \ref{sex4} we discuss the local stability of the black hole solution. In Section \ref{sex5} we discuss the phase transitions of the hairy black hole solution. Finally in Section \ref{sex6} we present our conclusions.

\section{The model}\label{sex2}

In the general setup, the total action is a contribution of the Einstein-Hilbert action $S_{EH}[g_{\mu\nu}]$, the action $S_{\p}[g_{\mu\nu},\p]$ for the scalar field $\p$ and the action $S_{M}[\hg_{\mu\nu},\psi]$ for the matter field $\psi$. The matter field is coupled to the so called matter metric $\hg_{\mu\nu}$ which is related to $g_{\mu\nu}$ through the relation $\hg_{\mu\nu}=g_{\mu\nu}+B[\p]\partial_{\mu}\p\partial_{\nu}\p$. Then, the total action corresponds to gravity minimally coupled to a scalar field and an electromagnetic field minimally coupled to the matter metric $\hg_{\mu\nu}$. Namely,
\bea\label{action}
I&=&\int d^4x \sdg \left(\frac{R-2\Lambda}{2\kappa}-\frac{1}{2}g^{\mu\nu}\na_{\mu}\p\na_{\nu}\p-\frac{1}{4} \hat{g}^{\mu \nu} \hat{g}^{\alpha \beta} F_{\mu \alpha} F_{\nu \beta}\right)\ ,
\eea
where $F_{\mu \nu}=\partial_{\mu} A_{\nu}-\partial_{\nu} A_{\mu}$ and $\hat{F}^{\alpha \beta}=\hat{g}^{\alpha \mu} \hat{g}^{\beta \nu} F_{\mu \nu}$. We obtain the following field equations,
\bea
\mathcal{E}^{\m\n}:=G^{\mu\nu}+\Lambda g^{\mu\nu}&=&\kappa(T^{\mu\nu}_{\p}+s\hat{T}^{\mu\nu})\ ,\label{geq}\\
\Box\p-B s \hat{T}^{\mu\nu}\hat{\na}_{\mu}\hat{\na}_{\nu}\p&=&0\ ,\label{KGs}\\
\hat{\nabla}_{\beta} \hat{F}^{\alpha \beta}&=&0\ .\label{meq}
\eea
with $s=\sdgb/\sdg$. The energy-momentum tensors are,
\bea
T^{\mu\nu}&=&-\frac{1}{2}g^{\mu\nu}(\na\p)^2+\na^{\mu}\p\na^{\nu}\p\ ,\\
\hat{T}^{\alpha \beta}&=&\hat{F}^{\alpha \mu} \hat{F}_{\mu}^{\beta}-\frac{1}{4} \hat{F}^{2} \hat{g}^{\alpha \beta}\ ,
\eea
where\footnote{Remind that $\hat{\nabla}^{\mu}\p=K\na^{\mu}\p$ and $\hat{\nabla}_{\mu}\p=\na_{\mu}\p$ with $K=1-B(\hat{\na}\p)^2$} $(\na\p)^2:=\nabla^{\mu}\p\nabla_{\mu}\p$, $(\hat{\na}\p)^2:=\hat{\nabla}^{\mu}\p\hat{\nabla}_{\mu}\p$, $\Box:=\nabla^{\mu}\nabla_{\mu}$, $\hat{\Box}:=\hat{\nabla}^{\mu}\hat{\nabla}_{\mu}$.

To find black hole solutions, it is convenient and simple to choose a line element with a radial coordinate such that $g_{tt}=-1/g_{rr}$ and a generic base manifold. Accordingly, we choose the following ansatz:
\begin{equation}\label{sol}
\begin{aligned}
ds^2&=-F(r)dt^2+\frac{dr^2}{F(r)}+r^2 d\Omega_\gamma^2\ ,\\
\phi&=\phi(r)\ ,\mathcal{A}=A(r)dt\ ,
\end{aligned}
\end{equation}
as well as the disformal function $B[\p]=B$ as a constant parameter, where $d\Omega^2$ stands for the line element of a 2-dimensional Euclidean manifold of constant curvature normalized to $\gamma=\{-1,0,1\}$, corresponding to a static hyperbolic, flat or spherical space, respectively.  In general, the field equations are quite complicated since the scalar field is present in every term where the matter metric is involved. However, following the same line of thought used in \cite{Hui:2012qt} to evade the no-hair theorems' obstructions, this technical but important detail can be significantly simplified, allowing us to obtain analytical solutions. This theory is shift symmetric under $\phi\to\phi+c$, and in consequence, the scalar field equation can be viewed as a conservation law, $\nabla_{\mu}J^{\mu}=0$, where the conserved current is given by,
\begin{eqnarray}
J^{\mu}&=&(g^{\mu\nu}-Bs\hat{T}^{\mu\nu})\nabla_{\nu}\p\ .\label{JU}
\end{eqnarray}
Note that the assumption of a regular horizon with a finite scalar current must have a vanishing radial component. This can be seen by demanding a finite norm of the scalar current $J^{\mu}J_{\mu}=(J^{r})^2/F$ at the horizon, which implies that $J^r$ must vanish there. According to this ansatz, the scalar field equation admits a first integral, $r^2 J^r=C_0$, where $C_0$ is an integration constant. Then, $J^r$ must vanish everywhere to satisfy this equation. This means,
\begin{equation}\label{seq}
J^{r}=F \psi\left[1+\frac{B\left(A^{\prime}\right)^{2}}{2\left(1+B F\psi^2\right)^{3 / 2}}\right]=0\ ,
\end{equation}
where $\psi=\phi'(r)$ and prime stands for derivation with respect to the radial coordinate. Note that $J^r$ is the only non-vanishing component of the current in \eqref{JU}. We immediately see a restriction to the parameter $B$, which must be negative to satisfy this relation. Using equation \eqref{seq}, we can solve the remaining Einstein equations, and as it is expected, the Maxwell equations are automatically solved by virtue of the Bianchi identity. It is found that the theory admits a charged hairy black hole solution (hBH). Namely,
\begin{equation}\label{sol}
\begin{aligned}
F(r)&=-\frac{r^2}{3}\left(\Lambda-\frac{\kappa}{2B}\right)+\gamma-\frac{2M}{r}-\frac{\kappa B q^4 }{40 r^6}\\
\psi^2(r)&=-\frac{1}{BF}\left(1-\frac{ B^2 q^4}{4r^8}\right)\ ,
A(r)=-\frac{B q^3}{10 r^5}\ ,
\end{aligned}
\end{equation}
where $M$ and $q$ are integration constants. We can see that this spacetime is asymptotically flat, dS or AdS when $\Lambda=\kappa/2B$, $\Lambda>\kappa/2B$ or $\Lambda<\kappa/2B$, respectively. In fact, identifying the effective cosmological constant $\Lambda_{eff}=\Lambda-\kappa/2B$, when the radial coordinate approaches infinity, the Riemann tensor takes the form
\begin{equation}
{R^{\alpha\beta}}_{\mu\nu}=\frac{\Lambda_{eff}}{3}{\delta^{\alpha\beta}}_{\mu\nu}\ ,
\end{equation}
while the spacetime solution,
\begin{equation}
g_{tt}=-g^{rr}\underset{r\rightarrow\infty}{\sim}\frac{\Lambda_{eff}}{3}r^2
\end{equation}
In the case of $\Lambda_{eff}\neq0$, the asymptotic behavior of $\psi^2$ goes like,
\begin{equation}\label{a1}
\psi^2(r)\underset{r \rightarrow \infty}{\sim}\frac{3}{B\Lambda_{eff} r^2}+O(r^{-4})\ ,
\end{equation}
and in the case of $\Lambda_{eff}=0$,
\begin{equation}\label{a2}
\psi^2(r)\underset{r \rightarrow \infty}{\sim}-\frac{1}{B\gamma}+O(r^{-1})\ .
\end{equation}
From \eqref{a1} and \eqref{a2} we see that there is a relation between a real scalar field and the asymptotic behavior. When $\Lambda_{eff}\neq0$, imposing the reality condition on the scalar field restricts the effective cosmological constant to $\Lambda_{eff}<0$, whereas in the asymptotically flat case $\Lambda_{eff}=0$ this condition is only compatible with a spherical base manifold. On the other hand, near the horizon $r_{+}$ we have,
\be
\psi^2(r)\underset{r \rightarrow r_{+}}{\sim}-\frac{1}{BF'(r_{+})(r-r_{+})}\left(1-\frac{ B^2 q^4}{4r_{+}^8}\right)+O(1)\implies\phi(r)\underset{r \rightarrow r_{+}}{\sim}2\sqrt{\frac{1}{-BF'(r_{+})}\left(1-\frac{B^2 q^4}{4r_{+}^8}\right)}\sqrt{r-r_{+}}+O(r-r_+)\ .
\ee
The black hole condition on the metric function $F$ at the horizon ensures the reality of the scalar field in the domain of outer communication provided,
\be
\frac{B^2 q^4}{4r_{+}^8}\leq 1\ .
\ee

A curvature singularity at the origin is dressed by the event horizon as it is evident from the Kretschmann scalar,
\begin{equation}
K=\frac{239 B^{2} q^{8} \kappa^{2}}{200 r^{16}}+\frac{56 B q^{4} M \kappa}{5 r^{11}}+\frac{2 B q^{4} \kappa \Lambda_{eff}}{3 r^{8}}+\frac{48 M^{2}}{r^{6}}+\frac{8 \Lambda_{eff}^{2}}{3}.
\end{equation}
It is worth mentioning that, there is no limit to a hairy neutral solution. It is because switching off the integration constant $q$, which we will see is the electric charge density, recovers the Schwarzschild black hole solution with a trivial scalar field. However, a second branch of solutions corresponds to the Reissner-Nordstr\"om AdS (RNAdS) black hole with a trivial scalar field $\phi=0$. We give more detail on this in the next section.

Up to here, we have just mentioned the event horizon located at $r_{+}$. Event horizons are located at $F(r)=0$, and since the metric function can be written as an eighth-degree polynomial,
\be
F=\frac{1}{120Br^6}P(r;M,q,B,\gamma)=\frac{1}{120Br^6}(-3 B^{2} q^{4}-240 B M r^{5}+12 \theta B r^{6} \gamma+20r^{8}(\kappa-2 B \Lambda))\ ,
\ee
there are eight roots: two real and six complex. The real roots represent the inner and outer horizon, $r_{-}$ and $r_+$, respectively.

As in the next section, any discussion of equilibrium black hole thermodynamics based on the Euclidean approach requires determining the allowed values of the parameters to define thermodynamic quantities at the boundary without any obstruction. It can be done by finding the extremality condition, where both event horizons coincide. For this, we study the discriminant of polynomial $P(r;M,q,B)$, which is a quite involved expression and not instructive to show it. Instead, we plot the results for each topology as follows,
\begin{figure}[H]
\centering
\includegraphics[width=.32\textwidth]{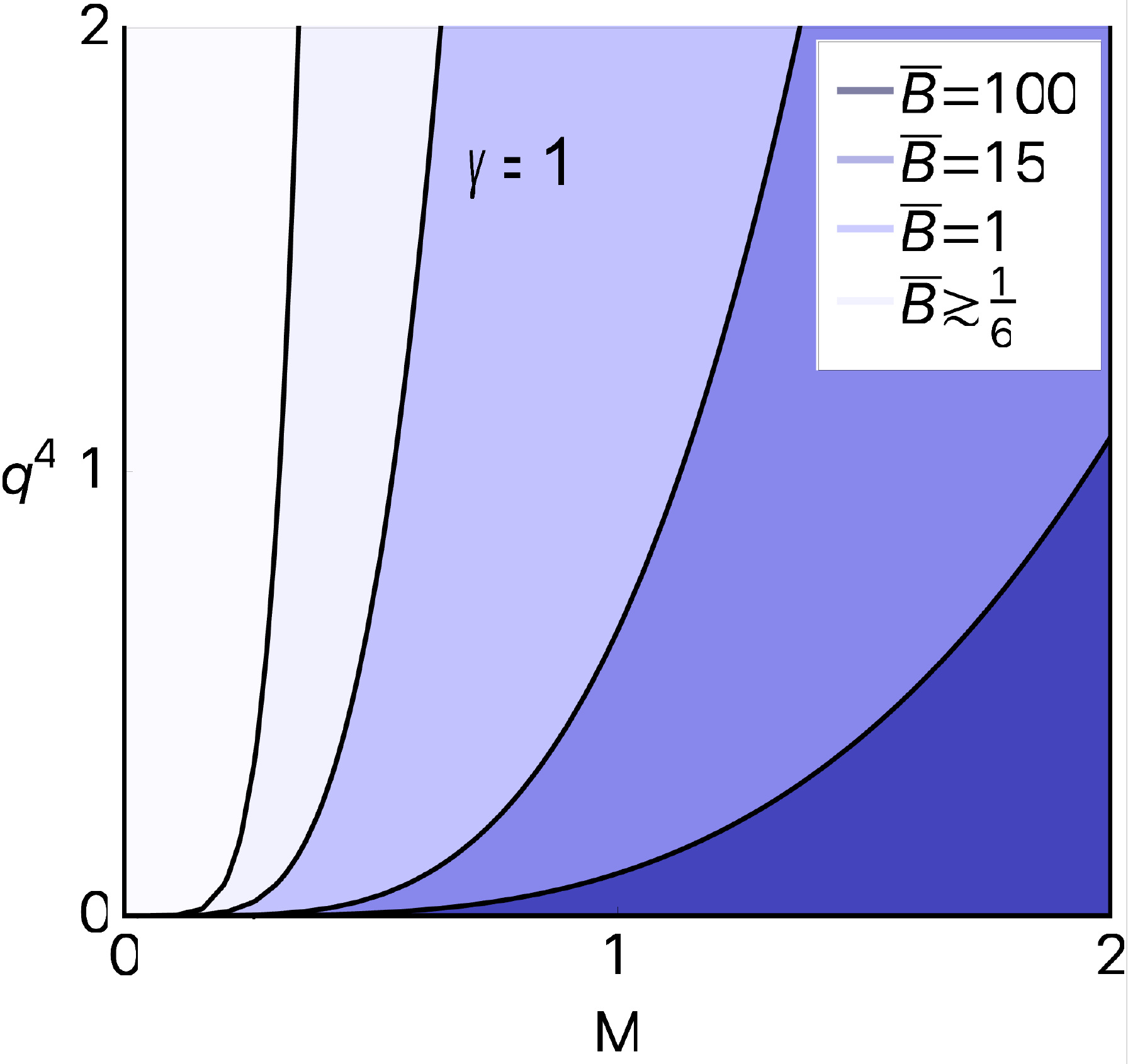}\hfill
\includegraphics[width=.32\textwidth]{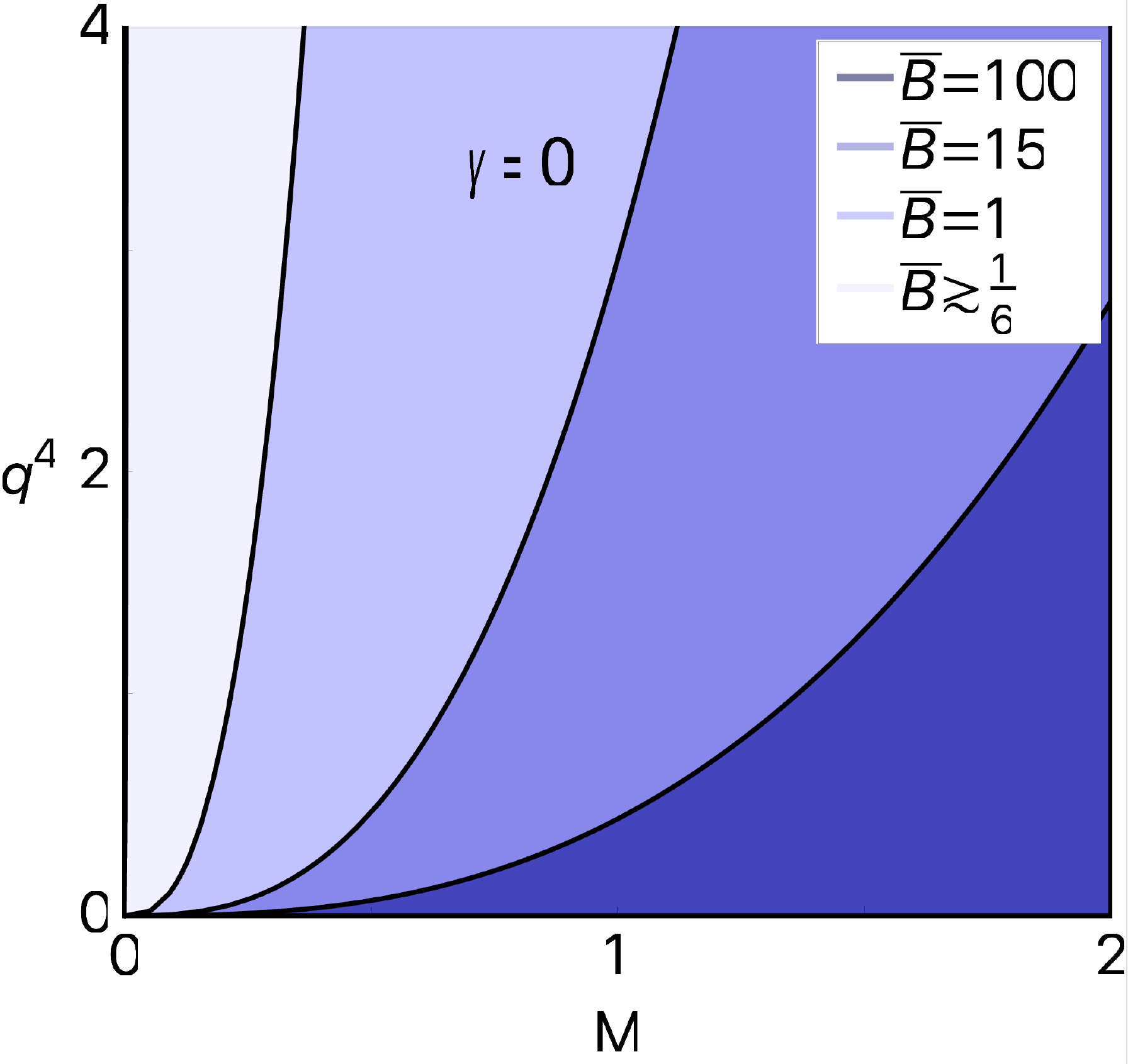}\hfill
\includegraphics[width=.335\textwidth]{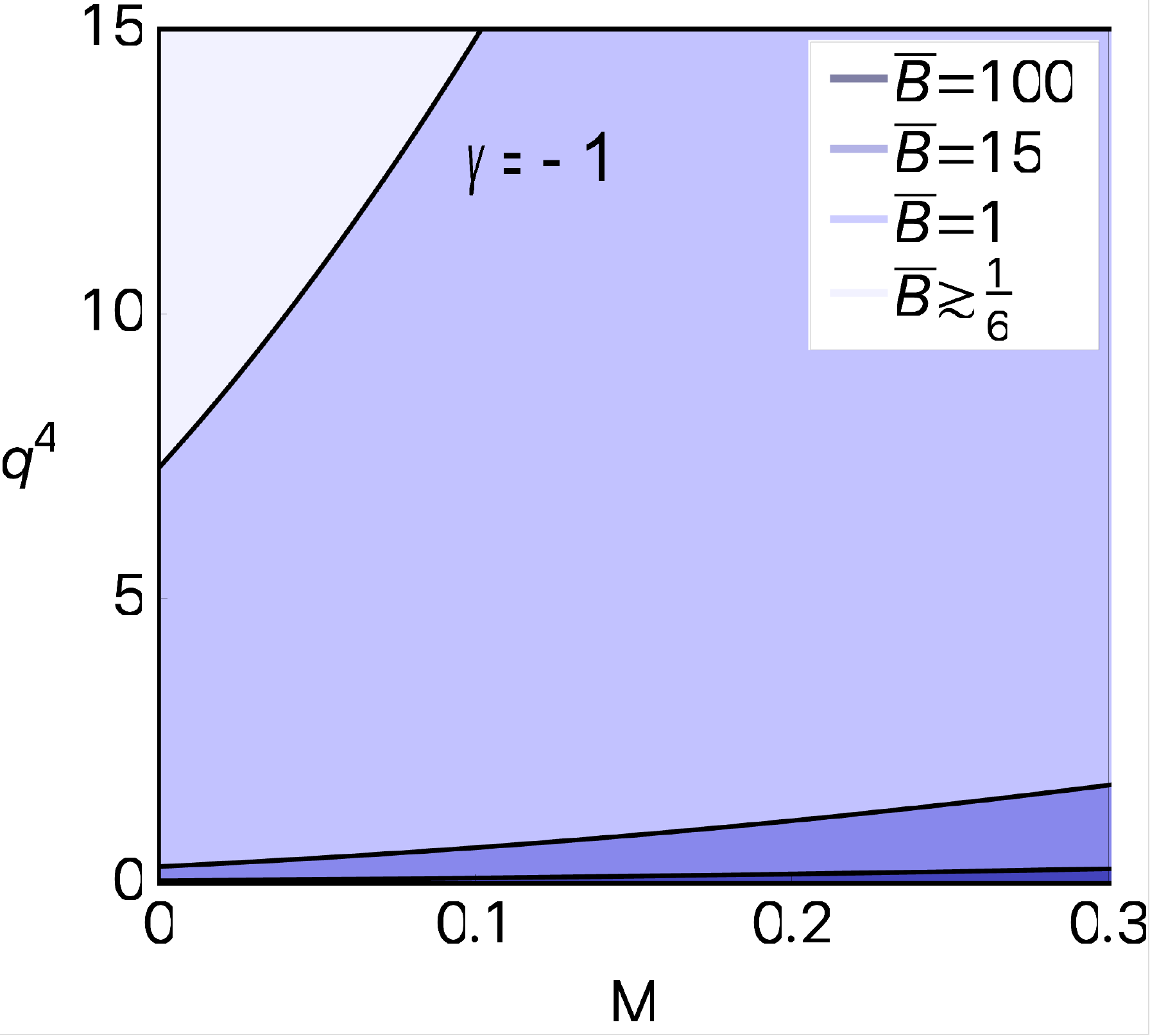}
\caption{Allowed values of $q^4$ for a given value of $M$ for different values of $\bb$. The whole region below the curves that separates them, represents the values of the mass where the hairy black hole possesses two event horizons.}
\label{ts}
\end{figure}
The plots depict the allowed region for the integration constant $q$ and $M$, which we will see are related to the conserved charges of mass and electric charge. Each region is delimited by the critical curve where $r_{-}=r_+$. The allowed values are in the whole region below the curve for some values of the parameter $B$.  Hereafter, to not carry negative signs in the values for $B$, we define $\bar{B}\equiv|B|$. We see that as we approach the asymptotically flat limit $\bb\sim1/6$ the set of allowed values drastically increases.

\section{Thermodynamics}\label{sex3}

In this section, we perform the thermodynamical analysis using the Euclidean approach. To achieve this, the partition function for a thermodynamical ensemble is identified with the Euclidean path integral in the saddlepoint approximation around the classical Euclidean solution \cite{Hawking:1982dh}.

The Euclidean continuation of the black hole configuration \eqref{sol} reads,
\begin{equation}\label{mini}
\begin{aligned}
d s_{E}^{2}&=N^{2}(r) F(r) d \tau^{2}+\frac{d r^{2}}{F(r)}+r^{2}d\Omega_\gamma^2\ ,\\
\phi&=\phi(r)\ ,\mathcal{A}=A(r) d \tau\ ,
\end{aligned}
\end{equation}
with $0\leq\tau\leq\beta$ periodic and $r>r_{+}$. A regular Euclidean geometry at the horizon requires a period $\beta$ of the Euclidean time identified with the inverse temperature $T=\beta^{-1}=F'(r_+)/4\pi$. The reduced Hamiltonian action is obtained replacing \eqref{mini} in \eqref{action} and can be written as follows,
\be
I_{E}=\beta\sigma \int_{r_{+}}^{\infty} d r \left(N \mathcal{H}-A_{\tau} \mathcal{G}\right)+B\ ,
\ee
where $\sigma$ is the area of the base manifold $\Omega_\gamma$, $B_{E}$ is a boundary term and
\bea
\mathcal{H}&=&\frac {r^2} {\kappa}\left( \frac{F'}{r}+\frac{F-\gamma}{r^2}+\Lambda +\frac {\kappa} {2}F\psi^2\right)+\frac {\left( \pi^{r}\right) ^{2}} {2r^{2}}\sqrt {1+BF\psi^2}\ ,\\
\mathcal{G}&=&\partial_{r}\pi^{r}\ .
\eea
Here, $\pi^r$ stands for the only non-vanishing component of the electromagnetic field momentum, defined by\footnote{These quantities can be obtained as a result of the standard Hamiltonian formalism applied to this theory.},
\begin{equation}
\pi^{r}=-\frac{r^{2} A^{\prime}}{N\sqrt {1+BF\psi^2}}\ .
\end{equation}
The boundary term $B_E$ is fixed by requiring that the action attains an extremum in the class of configuration considered. The variations on the reduced action respect to $N, F, A, \pi^r, \phi$ provides the following equations of motion,
\begin{eqnarray}
\mathcal{H}=0\ ,\label{H}\\
-\frac{r N^{\prime}}{\kappa}+\frac{1}{2} r^{2} N \psi^{2}+\frac{B N\left(\pi^{r}\right)^{2} \psi^{2}}{4 r^{2} \sqrt{1+B F \psi^{2}}}=0\ ,\label{varF}\\
\mathcal{G}=0\ ,\label{GL}\\
A'+\frac{N\pi^r\sqrt{1+BF\psi^2}}{r^2}=0\ ,\label{varpi}\\
\partial_r(N r^2 J^r)=0\ \label{varphi}\ ,
\end{eqnarray}
respectively. Here, $J^r$ is the conserved current already defined. It can be shown that these equations are consistent with the Einstein equations i.e. the black hole solution is also a solution of this set of equations. In fact, equation \eqref{varpi} is nothing else than the definition of the electromagnetic momentum. Equation \eqref{GL} is solved by $\pi^r=q$, with $q$ an integration constant. Then, from \eqref{varphi}, we obtain $J^{r}=j$, whose integration constant $j$ can be arbitrarily taken as zero, in order to have analytical solutions. Finally, it is straightforward to show that those equations provides $\psi^2$ and $A$, which along \eqref{H} and \eqref{varF}, determines $N$ as a constant that, without loss of generality, can be taken as $N=1$, and also $F$, which is giving by \eqref{sol}.

The variation of the boundary term gives,
\be\label{deltaB}
\delta B=\beta\sigma \left[N\left(-\frac{r \delta F}{\kappa}-r^{2} J^{r} \delta \phi\right)+A \delta \pi^{r}\right]^{\infty}_{r_+}\ ,
\ee
where $r_+$ represents the outer event horizon. As we saw in the previous section, real scalar fields are compatible with AdS or flat asymptotics. Since AdS black holes admit any horizon topology we consider $\Lambda_{eff}<0$ through the rest of this work\footnote{Another motivation, is that these solutions allow for holographic applications as it discussed in the conclusions.}. For this black hole solution we get the following contributions from the fields evaluated at the event horizon and at infinity respectively,
\be
\begin{aligned}
\delta F|_{r_{+}}&=-\frac{4 \pi}{\beta} \delta r_{+}\ ,&\delta \phi|_{r_{+}}&=\delta \phi\left(r_{+}\right)-\phi^{\prime}|_{r_{+}} \delta r_{+}\ ,&\delta \pi^{r}|_{r_{+}}=\delta q\ ,\\
\delta F|_{\infty}&=-2 \frac{\delta M}{r}-\frac{B K \delta q^{4}}{40 r^{6}}\ ,&\delta \phi|_{\infty}&=\sqrt{\frac{3}{\beta}} \frac{\delta M}{\Lambda_{eff}^{3 / 2} r^{3}}\ ,+O(r^{-5})\ ,&\delta \pi^{r}|_{\infty}=\delta q\ .
\end{aligned}
\ee
Notice that due to the condition $J^r=0$, only the first and third term in \eqref{deltaB} contribute, getting,
\begin{eqnarray}
\delta B(r_+)&=&\delta\left(\frac{A_+}{4 G}\right)+\beta \Phi \delta(\sigma q)\ ,\label{deltaBr}\\
\delta B(\infty)&=&\beta\sigma\frac{2 \delta M}{\kappa}\ ,\label{deltaBinf}
\end{eqnarray}
where we have identified the chemical potential for the electric field as $\Phi=A(r_+)$ and used that $A_+=\sigma r_{+}^{2}$ is the horizon area. At this point, we adopt the grand canonical ensemble, where the temperature $T=\beta^{-1}$ and the chemical potential are fixed. Using the boundary conditions we integrate the variations of the boundary term obtaining the expressions,
\bea
B(r_+)&=&\frac{A_{+}}{4 G}+\beta \Phi \sigma q\ ,\\
B(\infty)&=&\beta\sigma\frac{2M}{\kappa}\ ,
\eea
and the value of the reduced action on-shell reads,
\be\label{Ie}
I_E=\beta\sigma\frac{2M}{\kappa}-\frac{A_{+}}{4 G}+\beta \Phi \sigma q\ ,
\ee
up to an arbitrary additive constant without variation. In the grand canonical ensemble, the Gibbs energy $\mathcal{G}$ is related to the Euclidean action by $I_E=\beta \mathcal{G}=\beta \mathcal{M}-S-\beta \Phi Q$, where the mass $\mathcal{M}$, the electric charge $Q$ and the entropy $S$, are computed as usual, obtaining,
\be\label{termosHBH}
\begin{aligned}
\mathcal{M}&=\left(\frac{\partial}{\partial \beta}-\beta^{-1} \Phi \frac{\partial}{\partial \Phi_{E}}\right)I_E=\sigma\frac{2M}{\kappa}\ ,&Q&=-\frac{1}{\beta} \frac{\partial I_{E}}{\partial \Phi_{E}}=\sigma q\ ,&S &=\left(\beta \frac{\partial}{\partial \beta}-1\right) I_{E}=\frac{A_{+}}{4 G}\ .
\end{aligned}
\ee
As a consequence of this approach, the first law of thermodynamics is satisfied $d \mathcal{M}=T d S+\Phi d Q$, which is the case considering the last expressions.

There is a second branch of black hole solution which is not smoothly connected to the previous one when the scalar field is switched off. This means that the thermodynamical system can admit a second configuration in the same grand canonical ensemble. As we mentioned in the previous section, this is nothing else than the RNAdS solution with trivial scalar field $\phi=0$, which reads,
\begin{equation}
\begin{aligned}
F_0(\r)&=-\frac{\r^{2}}{3} \Lambda+\gamma-\frac{2 M_0}{\r}+\frac{\kappa q_0^{2}}{2 \r^{2}}~,\\
\psi_0^2(\r)&=0\ ,
A_0(\r)=-\frac{q_0}{\r}\ .
\end{aligned}
\end{equation}
For the RNAdS black hole, the mass $\mathcal{M}_0$, the electric charge $Q_0$ and the entropy $S_0$ are given by,
\be
\begin{aligned}\label{termos}
\mathcal{M}_0&=\sigma\frac{2M_0}{\kappa} \ ,&Q_0&=\sigma q_{0}\ ,&S_0&=\frac{A_{+0}}{4 G}\ ,
\end{aligned}
\ee
where $\rho_+$ stands for the outer event horizon, $A_{+0}=\sigma \r_{+}^{2}$ is the horizon area and the chemical potential for the electric field can be identified as $\Phi_0=A_0(\rho_+)$.

At this stage, it is convenient to use $l$ as a length scale and rescale thermodynamic quantities to express them as dimensionless variables. Namely,
\be
\begin{aligned}
T&\rightarrow \frac{T}{l}&\mathcal{M}&\rightarrow \sigma l\mathcal{M}&S&\rightarrow \sigma l^2S\ ,\\
\Phi&\rightarrow\Phi&Q&\rightarrow\sigma lQ&\mathcal{G}&\rightarrow\sigma l\mathcal{G}\ .
\end{aligned}
\ee
By consistency with this rescaling we promote $r\rightarrow r l$, $q\rightarrow q l$, $M\rightarrow M l$ and the parameter $B\rightarrow B l^2$ to dimensionless variables. The same rescaling runs for the corresponding RNAdS quantities. With this rescaling, we get rid of carrying the parameter $l$ in any of the following equations. Throughout this work, we will use these variables and focus on the case of AdS asymptotics, which requires $\Lambda=-3/l^2$ and $\bar{B}>1/6$. This way, both black hole spacetimes share the same asymptotic behavior. We also set $\kappa=8\pi G=1$, unless $G$ appears explicitly.

\begin{figure}[H]
\centering
\includegraphics[width=.33\textwidth]{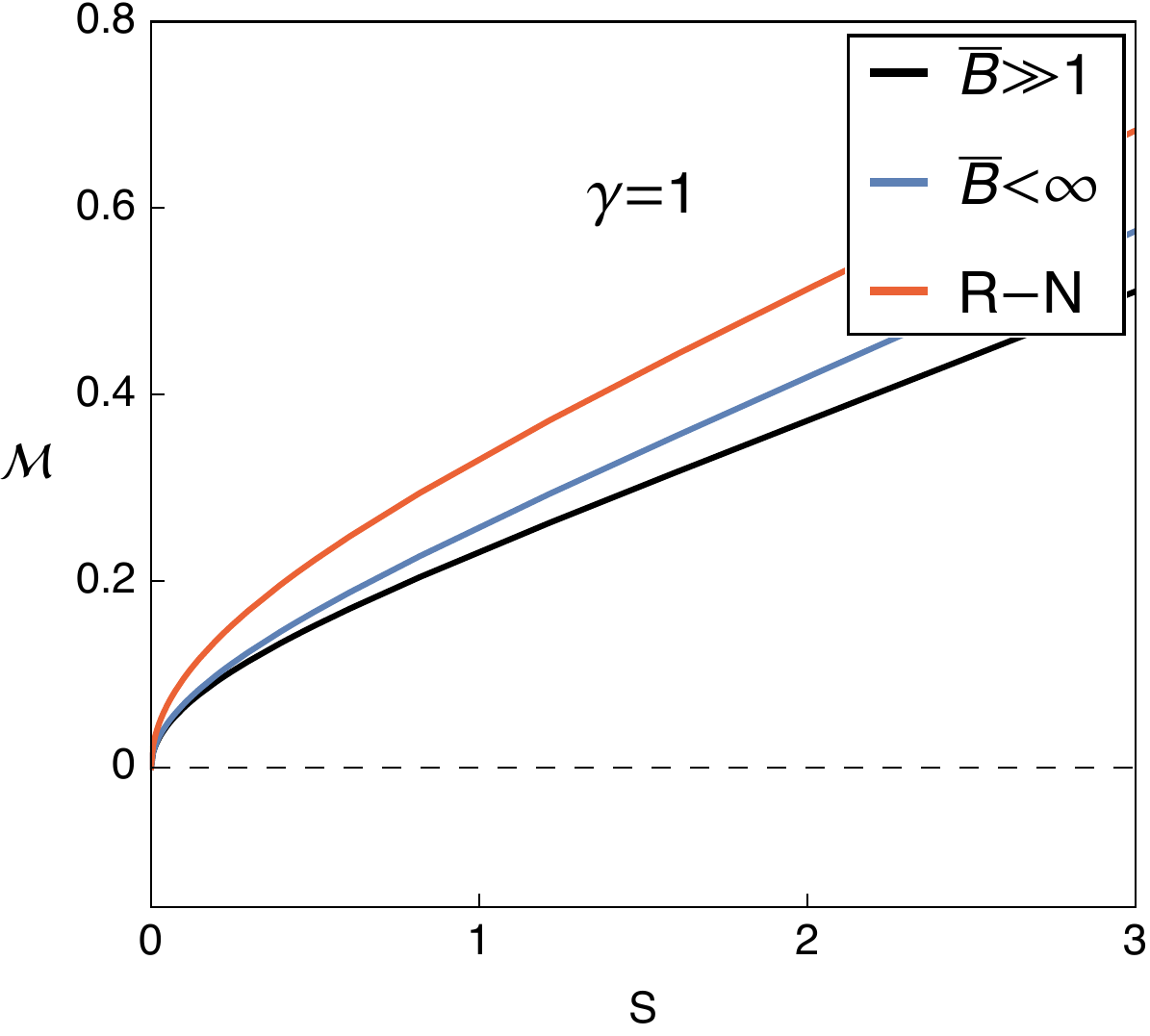}\hfill
\includegraphics[width=.33\textwidth]{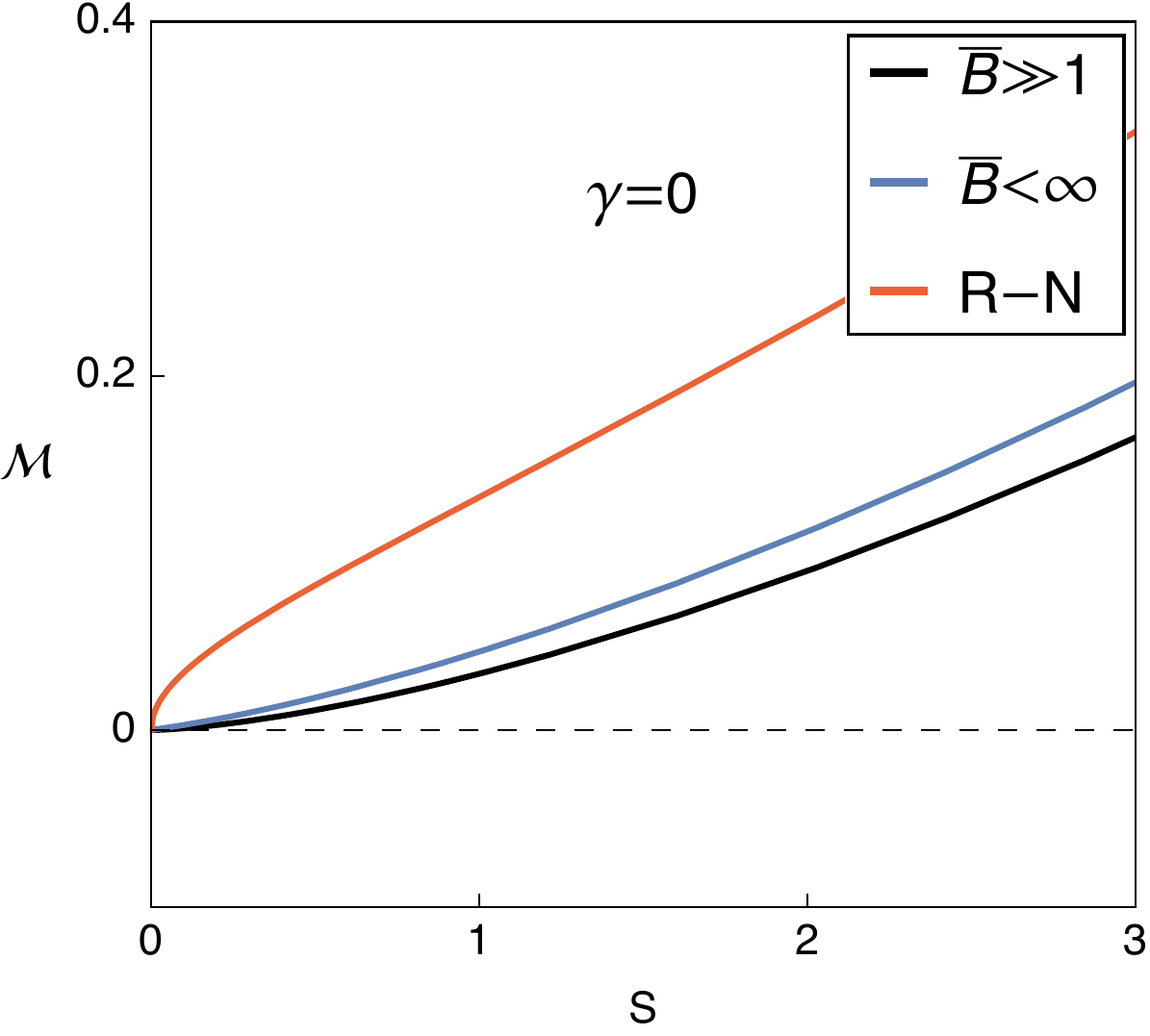}\hfill
\includegraphics[width=.33\textwidth]{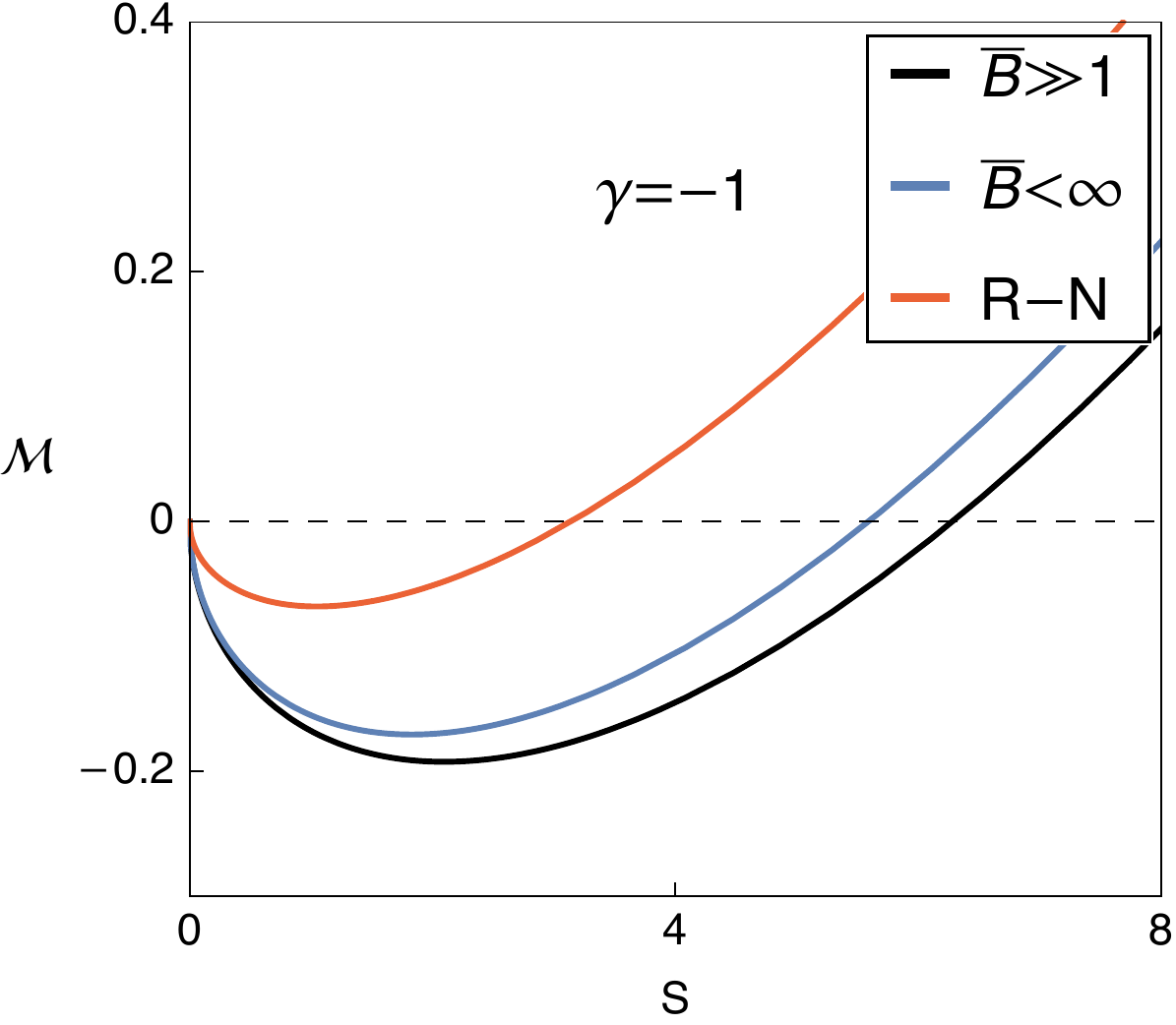}
\caption{Mass as a function of the entropy for fixed potential. For comparison we have included the red curve corresponding to Reissner-Nordstr\"om (R-N) black hole, while the blue curve is representative of the curves for any value of $\bb$. In particular, we used $\bb=10$. The electric potential has been set to $\Phi=1$.}
\label{ms}
\end{figure}

As expected, the mass of the black hole strongly depends on the coupling parameter $B$. It is given by,

\begin{equation}\label{mvss}
\mathcal{M}=\frac{\gamma \sqrt{S}}{2 \sqrt{2 \pi}}-\frac{(1-6 \bb) S^{3 / 2}}{24 \sqrt{2} \pi^{3 / 2} \bb}+\frac{5^{1 / 3} \Phi^{4 / 3} S^{5 / 6}}{8 \sqrt{2} \pi^{5 / 6} \bb^{1 / 3}}
\end{equation}

and plotted in Fig. \ref{ms} for a fixed potential. We have included the RNAdS mass for comparison.

We see that the hairy black hole with spherical and flat base manifold possesses positive mass irrespective of its size. In contrast, in the hyperbolic case, the mass is no longer a monotonically increasing function, and black holes with negative mass are allowed, which imposes a physical restriction on the allowed size of these black holes. For a given positive mass at the same electric potential, the entropy of the hBH is larger than the entropy of the RNAdS, even in the asymptotically flat case $\bar{B}=1/6$. The difference of the entropy increases as the coupling parameter increases, reaching a limit curve when the scalar field is strongly coupled given by $\M=\frac{\sqrt{S}(S+2 \pi \gamma)}{4 \sqrt{2} \pi^{3 / 2}}$, as it can be seen in the black curve and obtained from \eqref{mvss}. Although it is not conclusive, it strongly suggests that the hBH is most probably stable than its GR counterpart. We can also interpret from these curves that the strength of the coupling parameter reduces the black hole mass.

\section{Local stability}\label{sex4}

In this section, we analyze the local stability of the black hole solution by studying its response to the system under small perturbations of its thermodynamical variables around the equilibrium. There are many equivalent criteria in the literature, such as the sign of the second derivative of the entropy and the energy or any of its associated Legendre transforms. Since we are considering the grand canonical ensemble, here, we choose to analyze the local stability by computing the heat capacity at constant electric potential and electric permittivity at a constant temperature. This is, respectively,
\be\label{ls}
\begin{aligned}
C_{\Phi}&\equiv T\left(\frac{\partial S}{\partial T}\right)_{\Phi}\ ,&\epsilon_{T}&\equiv\left(\frac{\partial Q}{\partial \Phi}\right)_{T}\ .
\end{aligned}
\ee
To obtain this expression, we need the temperature against the entropy for a fixed potential. This expression is given by,
\be\label{temp}
T=\frac{\gamma}{2 \sqrt{2 \pi} \sqrt{S}}-\frac{5 \times 5^{1 / 3} \Phi^{4 / 3}}{8 \sqrt{2} \pi^{5 / 6}  \bar{B}^{1 / 3}S^{1 / 6}}-\frac{(1-6 \bb)\sqrt{S}}{8 \sqrt{2} \pi^{3 / 2} \bar{B}}\ ,
\ee
which can be seen in Fig. \ref{ts} as follows:
\begin{figure}[H]
\centering
\includegraphics[width=.33\textwidth]{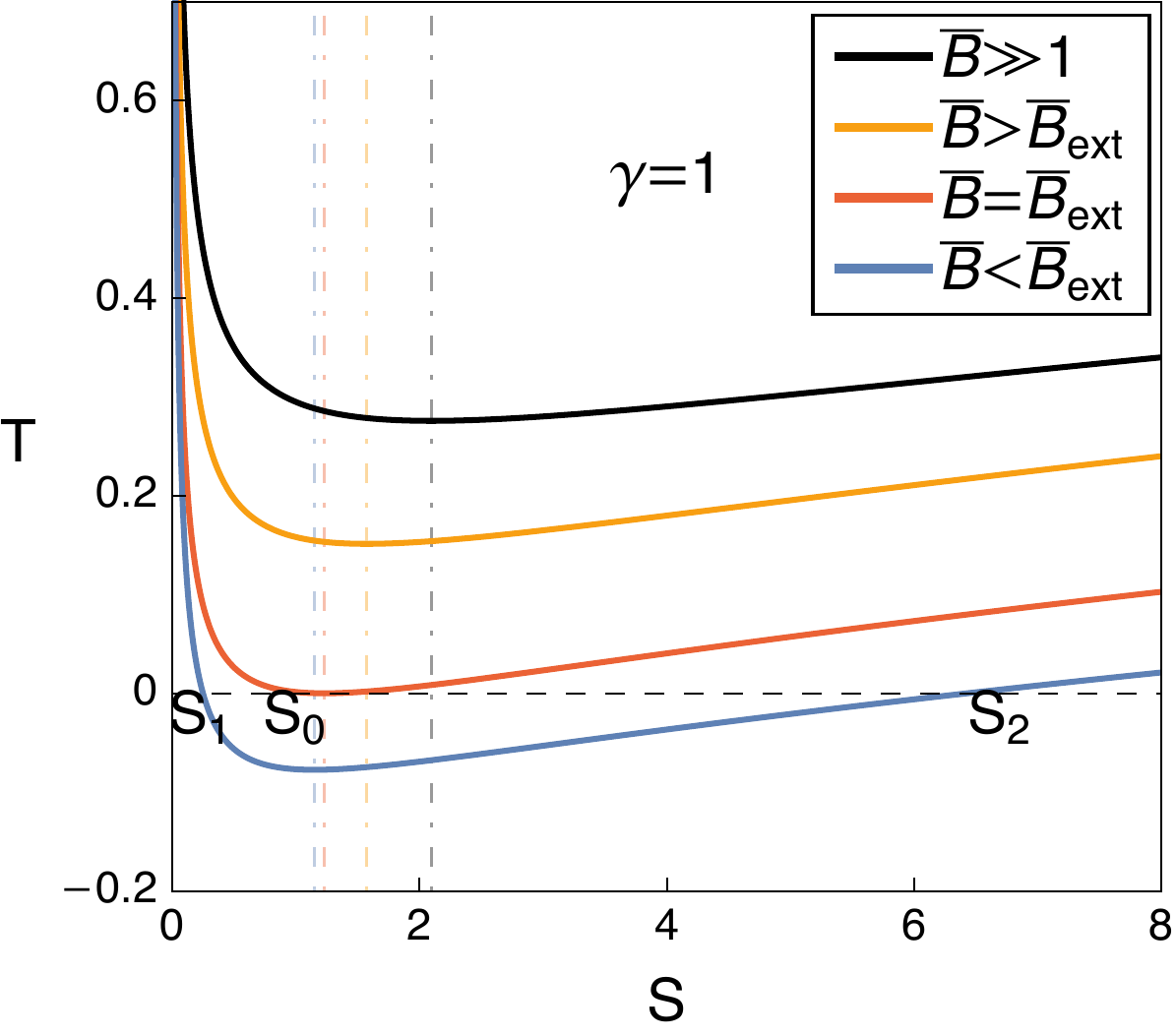}\hfill
\includegraphics[width=.33\textwidth]{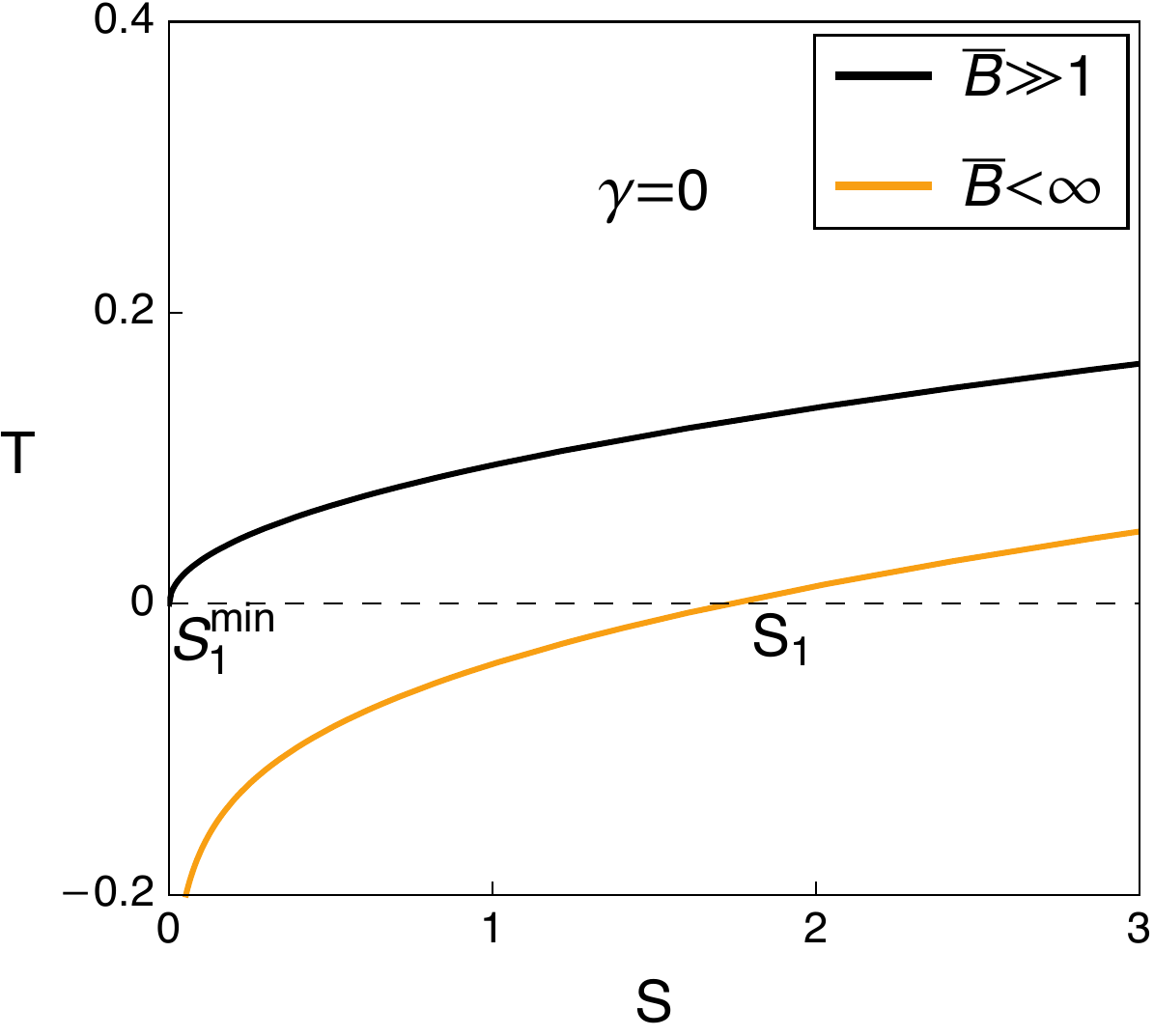}\hfill
\includegraphics[width=.33\textwidth]{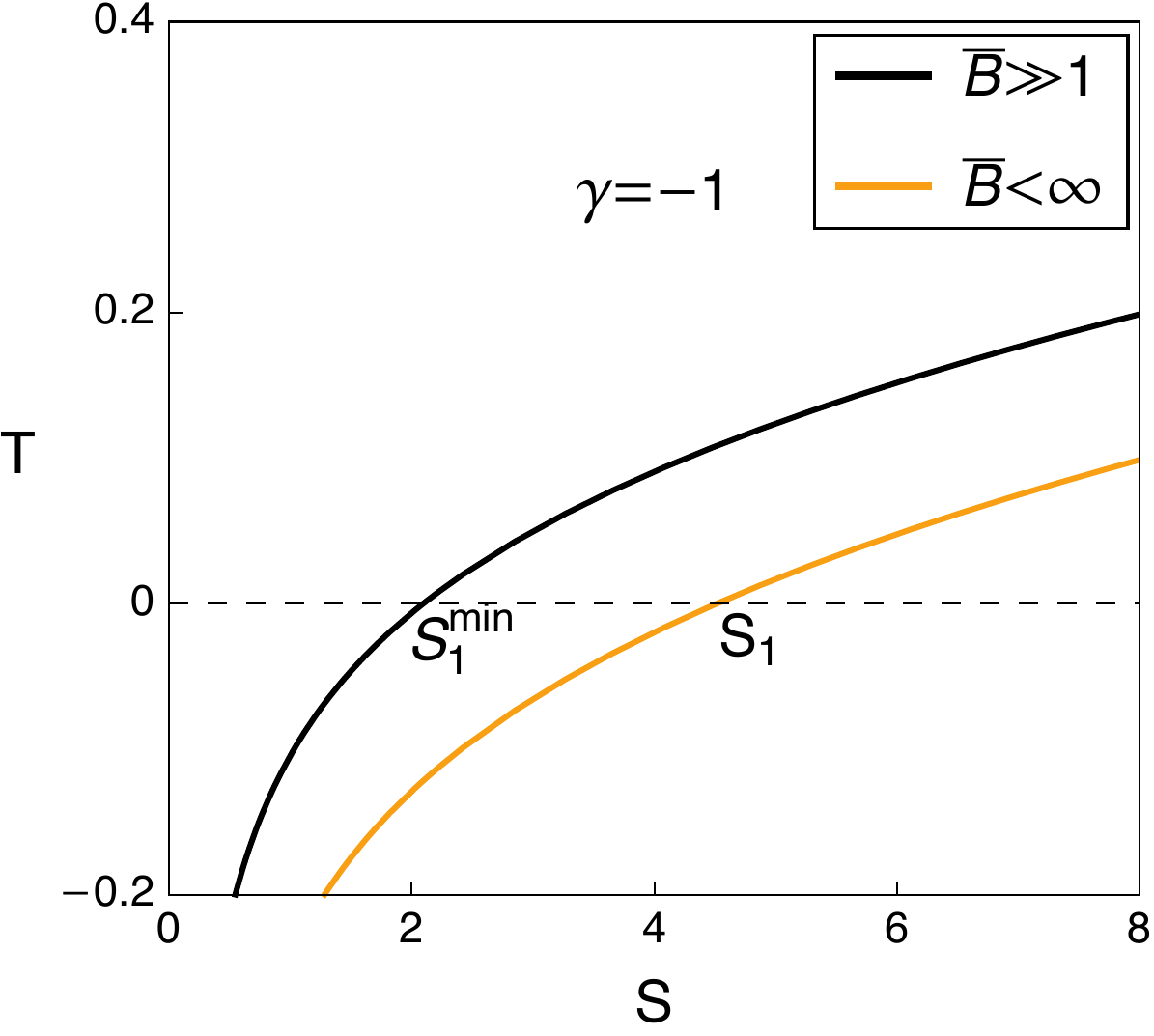}
\caption{Temperature as a function of the entropy for fixed potential. The electric potential has been set to $\Phi=1$ and the curves are representatives for each case, where have used in particular $\bb=0.6$ for $\bb<\bex$, $\bex=1.13$, $\bb=10$ for $\bb>\bex$ and $\bb<\infty$, and the analytic curve for $\bb\gg1$.}
\label{ts}
\end{figure}
In the spherical case, $S_0$ generically represents the value of entropy associated with the minimum temperature. For each curve the respective vertical dot-dashed line crosses the entropy axis at that particular value of the entropy. We see that below the extremal value of $\bb$ there are two extremal black holes with entropies $S_1$ and $S_2$, which belong to the small and large branch, respectively.

\begin{figure}[H]
\centering
\includegraphics[width=.33\textwidth]{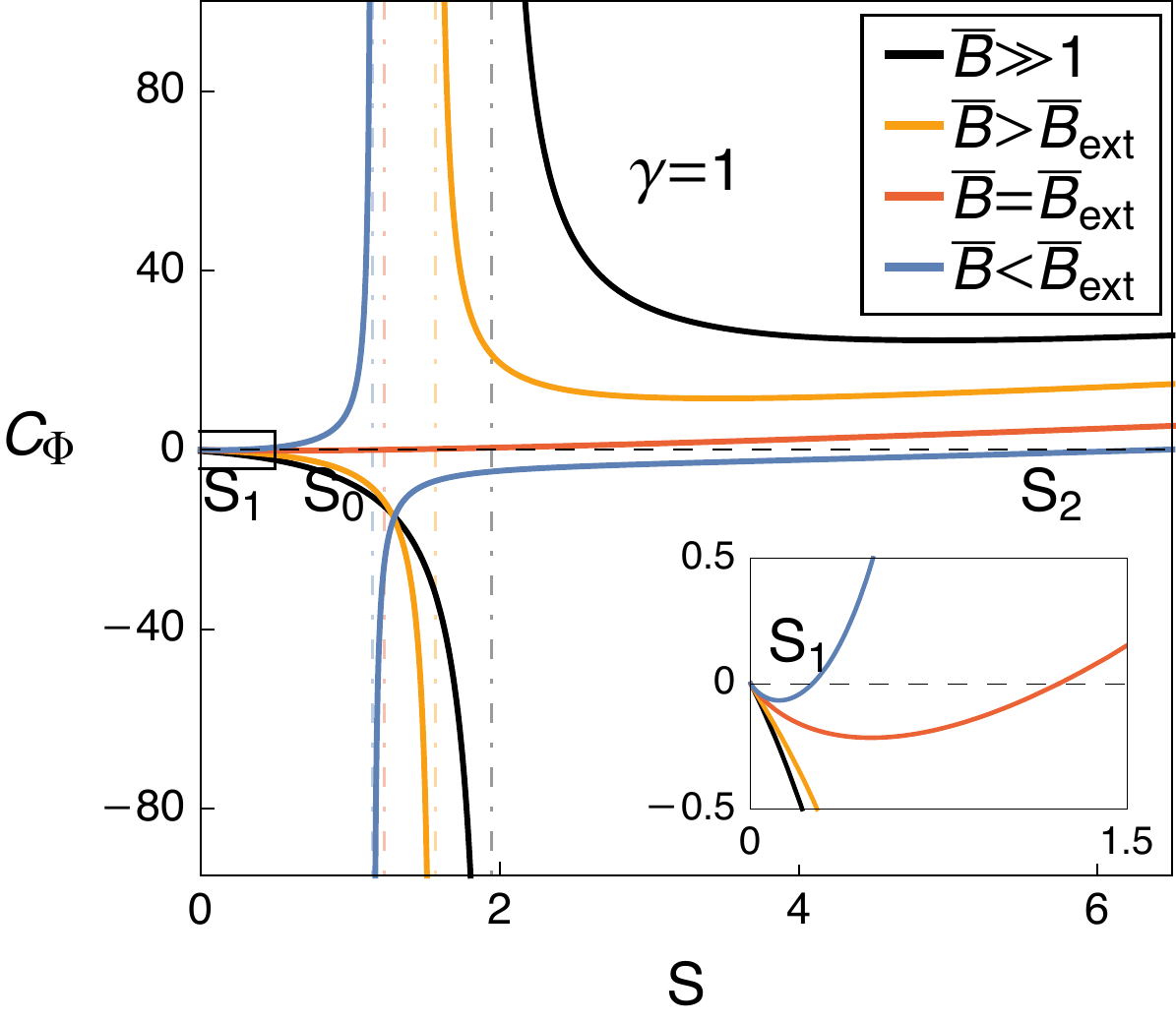}\hfill
\includegraphics[width=.33\textwidth]{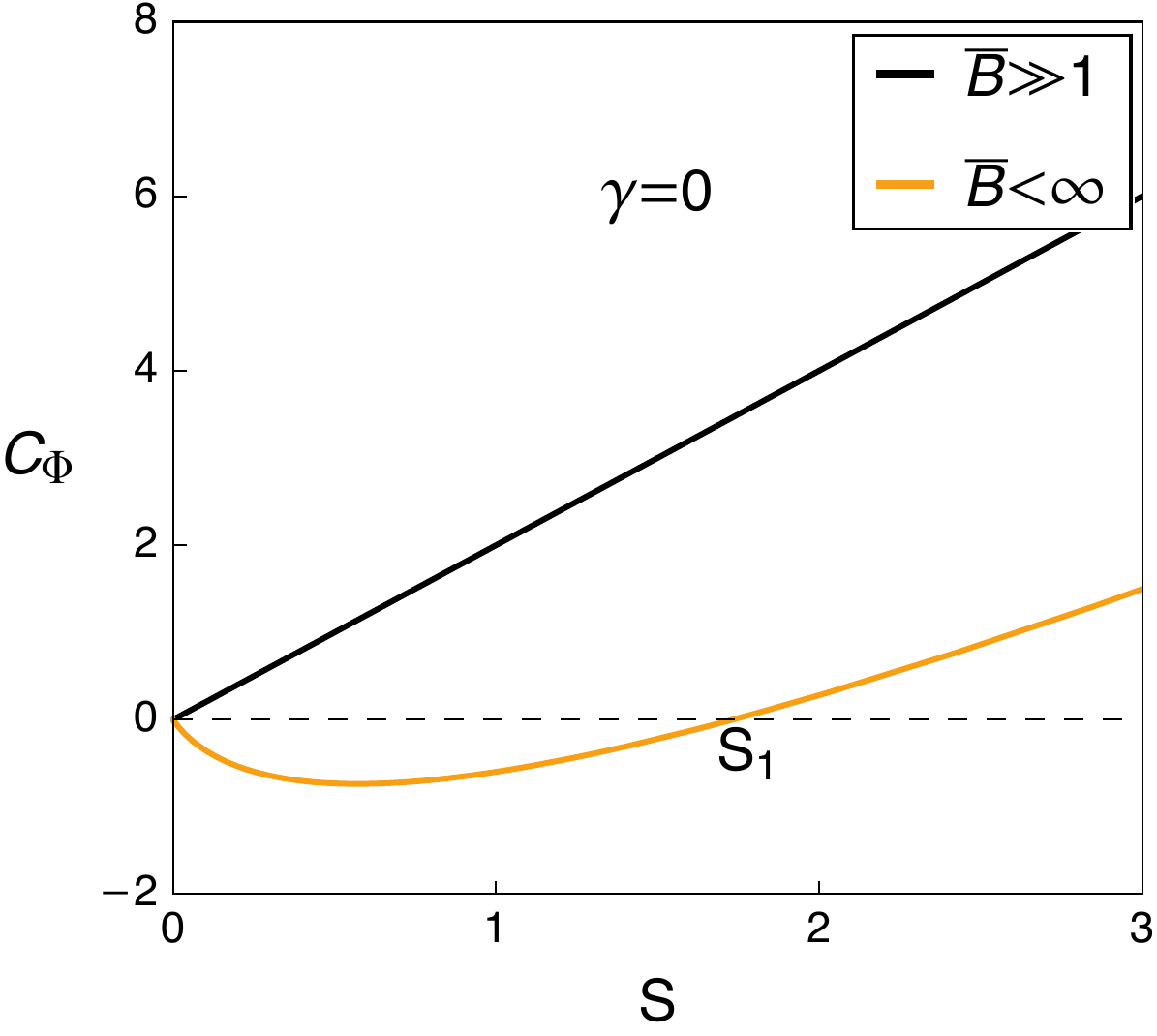}\hfill
\includegraphics[width=.33\textwidth]{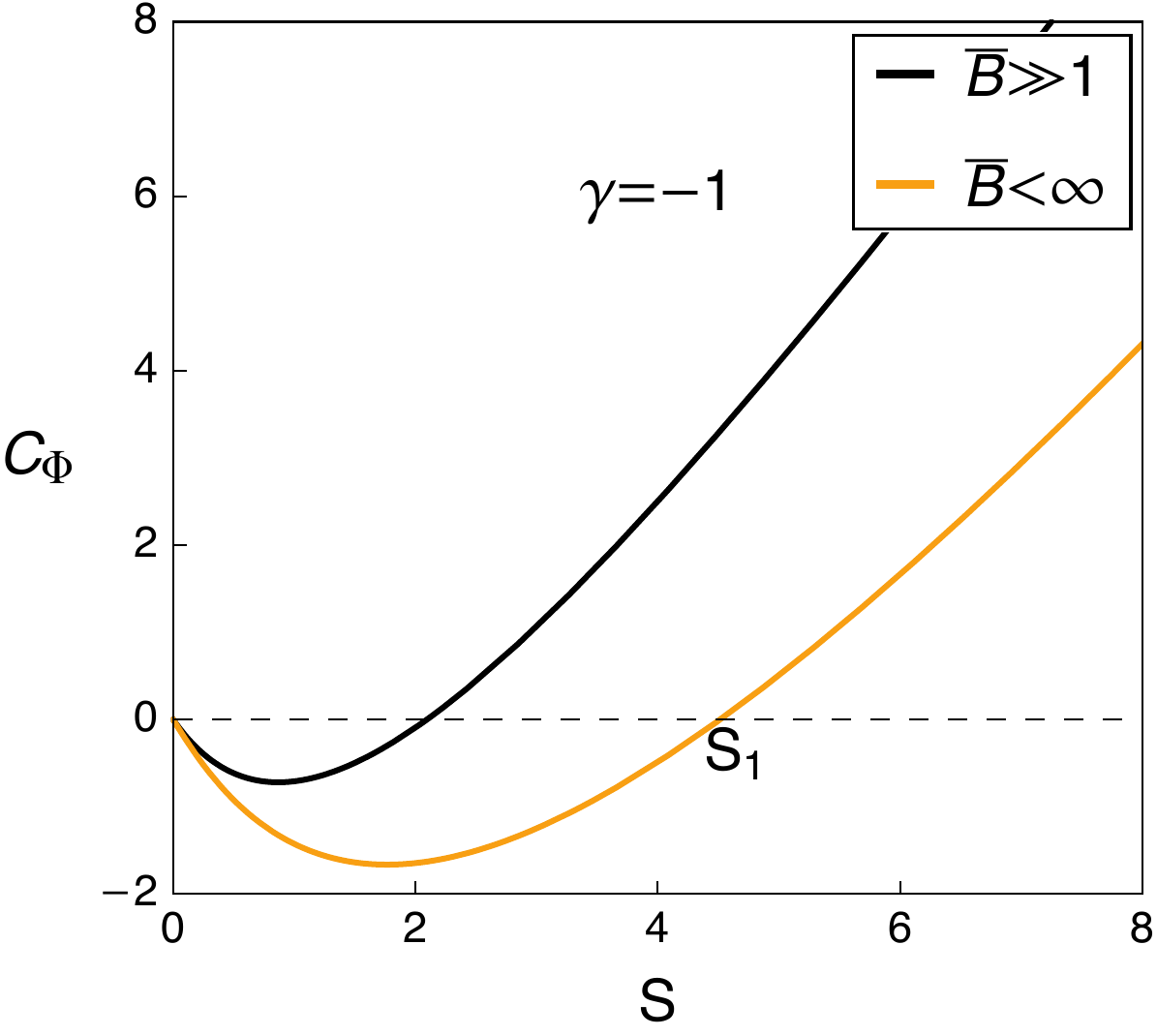}\\
\includegraphics[width=.33\textwidth]{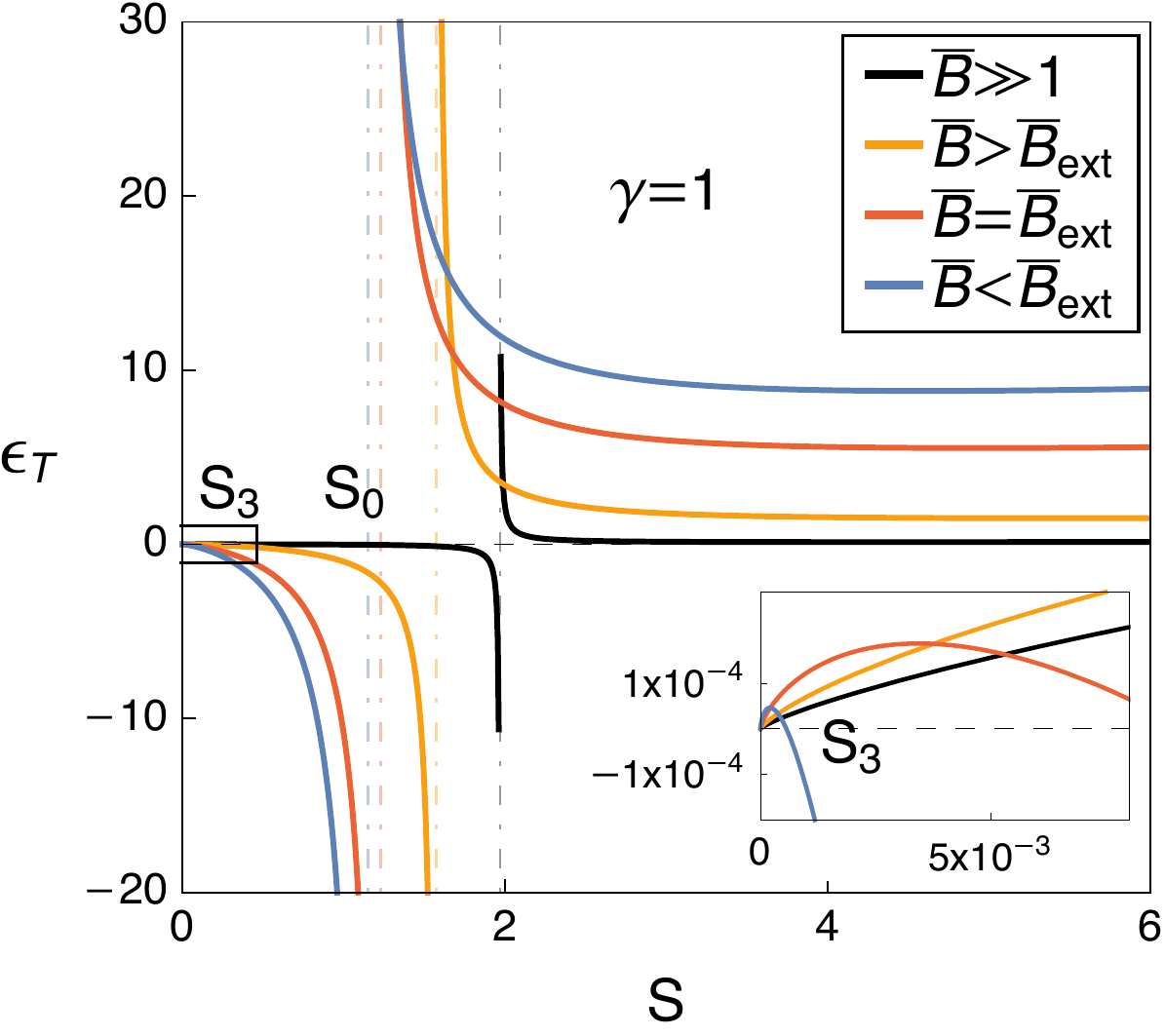}\hfill
\includegraphics[width=.33\textwidth]{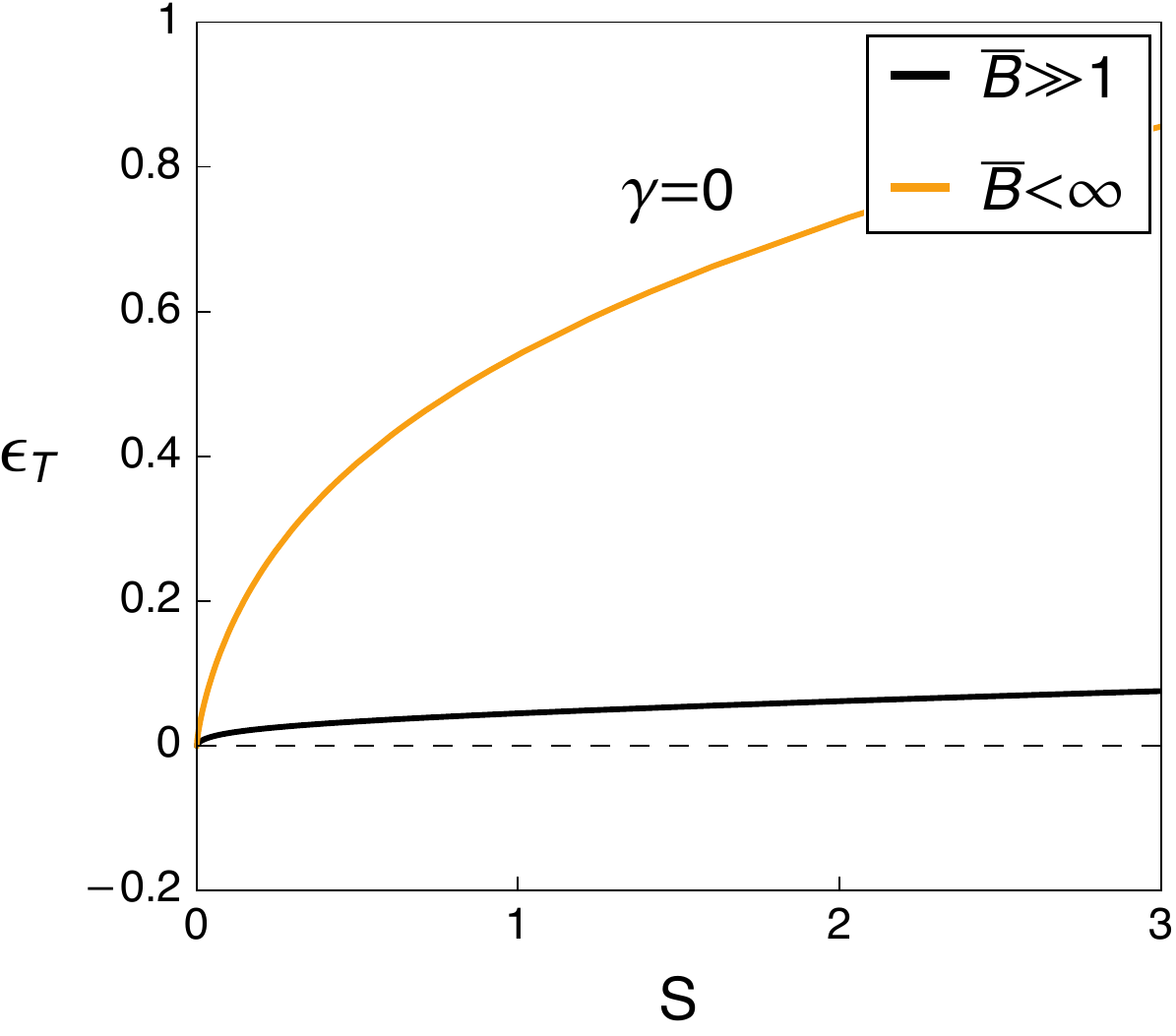}\hfill
\includegraphics[width=.33\textwidth]{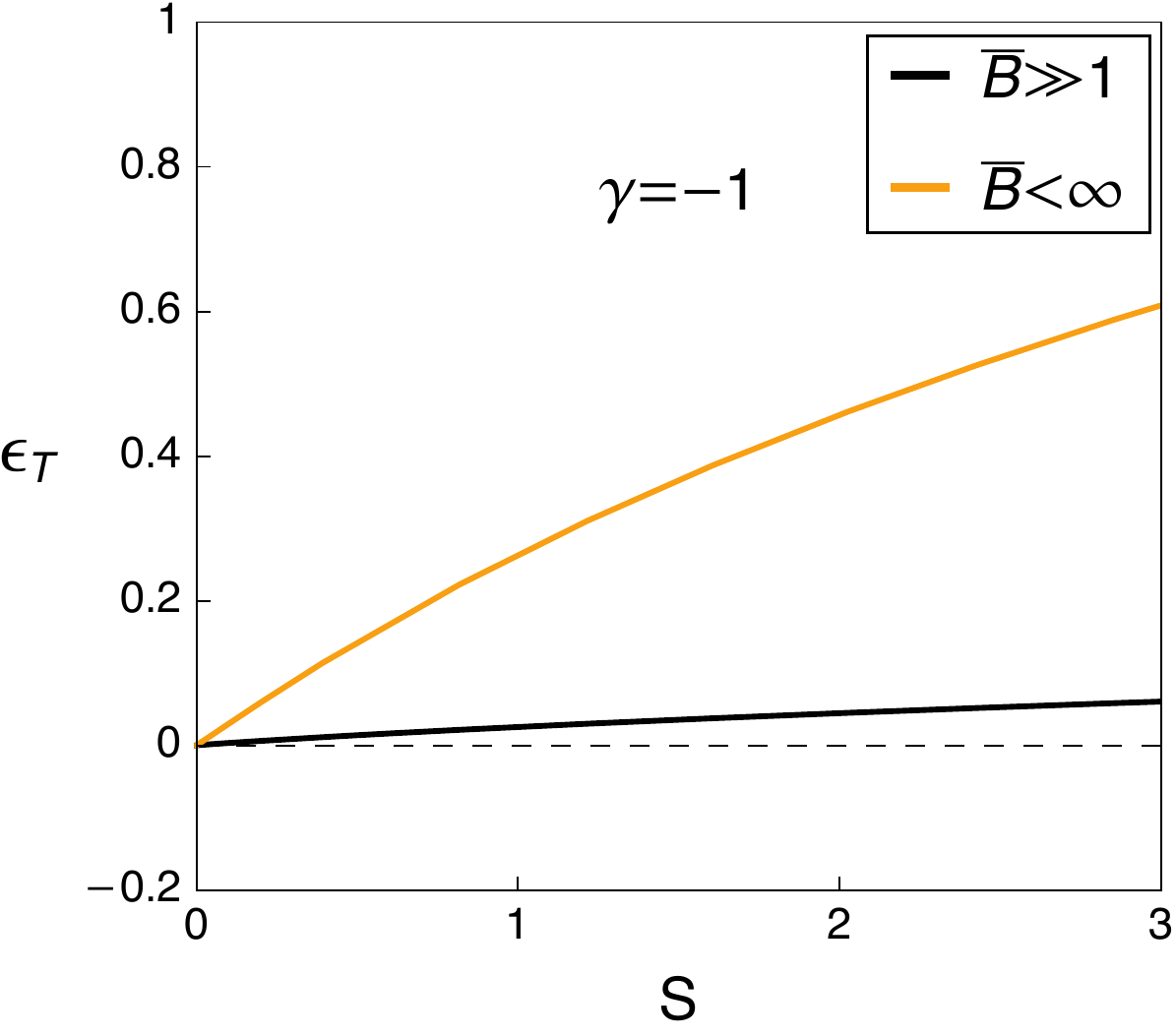}
\caption{Heat capacity in the first row and electric permittivity in the second row as a function of the entropy for fixed potential. We have used the same values for the electric potential and $\bb$ than in Fig. \ref{ts}.}
\label{ce}
\end{figure}

From \eqref{ls}, the heat capacity and electric permittivity are explicitly given by,
\bea
C_{\Phi}&=&\frac{6 S\left(-4 \pi \bar{B}\gamma+5 \times 5^{1 / 3} \pi^{2 / 3} \bar{B}^{2 / 3} \Phi^{4 / 3}S^{1 / 3}+(1-6 \bar{B})S\right)}{12 \pi \bar{B}\gamma-5 \times 5^{1 / 3} \pi^{2 / 3}  \Phi^{4 / 3} \bar{B}^{2 / 3}S^{1 / 3}+3(1-6 \bar{B})S}\ ,\\
\epsilon_T&=&\frac{5^{1 / 3} S^{5 / 6}\left(4 \pi \gamma \bar{B}-35 \times 5^{1 / 3} \pi^{2 / 3} \bar{B}^{2 / 3} \Phi^{4 / 3}S^{1 / 3} +(1-6 \bar{B})S\right)}{\sqrt{2} \pi^{5 / 6} \bar{B}^{1 / 3} \Phi^{2 / 3}\left(12 \pi \bar{B}\gamma-5 \times 5^{1 / 3} \pi^{2 / 3}  \Phi^{4 / 3} \bar{B}^{2 / 3}S^{1 / 3}+3(1-6 \bar{B})S\right)}\ .
\eea
The graphic results might be more enlightening than the analytical expressions, as can be seen in Fig. \ref{ce}.
In the strongly coupled regime, the $T-S$ curve takes a particularly simple form for any topology of the base manifold,
\be\label{tinf}
T_{\bb\gg1}=\frac{2 \pi\gamma+3 S}{4 \sqrt{2} \pi^{3 / 2} \sqrt{S}}\ .
\ee
\textbf{Case} {\boldmath$\gamma=1$}: In Fig. \ref{ce} we see that $T(S)$ reaches a global minimum $T_{\text{min}}$ at a value of the entropy which we generically denoted as $S_0$ for each curve. It corresponds to those values of the entropy where the dashed vertical lines cut the $S$ axis in the plot. This means that this value satisfies $\partial_{S}T(S_0)=0$ and is given by the relation
\be\label{s0}
12 \pi \bar{B}-5 \times 5^{1 / 3} \pi^{2 / 3}  \Phi^{4 / 3} \bar{B}^{2 / 3}S_0^{1 / 3}+3(1-6 \bar{B})S_0=0\ .
\ee
In consequence, the temperature $T_{\text{min}}$ corresponding to entropy $S=S_0$ can be obtained by using \eqref{temp} and \eqref{s0}, and represents the minimum temperature of the system. We can interpret this as there are two phases in the $T-S$ curve, separated by the point $(S_0,T_{\text{min}})$ in Fig. \ref{ts}. According to the expression for the entropy in \eqref{termos}, since it is a monotonic increasing function on $r_+$, there are two branches of black holes having small black holes for $S<S_0$ and large black holes for $S>S_0$, with both phases existing above the minimum temperature $T_{\text{min}}$. Unlike the RNAdS black hole, this solution admits two extremal black holes for a certain range of parameters, and it is always possible to find the two phases. We can divide the space of parameters $\bar{B}-\Phi$ in two regions, one with no extremal black holes and the other containing two extremal black holes, with a limit case where there is only one extremal black hole, i.e., $T_{\text{min}}=0$. The critical value $\bex$ represents the unique value at which the latter situation happens for a given potential. This can be found by casting equation \eqref{temp} to a cubic polynomial and analyzing the resultant discriminant. We find that,
\be\label{bext}
\bex=\frac{1}{6}+\frac{625 \Phi^{4}}{648}\ .
\ee
When the value of the coupling constant $B$ is such that $\bb<\bex$, there are values of the event horizon associated with entropies $S_1$ and $S_2$ which correspond to two extremal black holes. It means that $S_1$ and $S_2$ satisfy $T(S_1)=0=T(S_1)$ with $S_0\in [S_1,S_2]$ and $T_{\text{min}}<0$. Only when $\bb=\bex$, we get $S_0=S_1=S_2$. On the other hand, there is no extremal black hole for $\bb>\bex$ where $T_{\text{min}}>0$, and if we continue increasing the parameter $\bb$, in the strong coupled regime $T_{\text{min}}$ reaches its maximum possible value $T=\sqrt{3}/2\pi$ at $S_0=2\pi/3$ as we can see in Fig. \ref{ts} and obtain from \eqref{temp}. As $\bb$ decreases $S_0$ gets smaller until reach the limit case for asymptotically flat black holes when $\bb=1/6$ approaching to an entropy given by $S_0\rightarrow32\pi/(1875\Phi^4)$. In consequence, the value of $S_0\in(32\pi/(1875\Phi^4),2\pi/3]$ when we dial $\bb\in(1/6,\bb\gg1)$, as it can be seen in the sequence of vertical dashed lines. From \eqref{bext} we check that $\bex>1/6$, which ensures that all the electrically charged asymptotically AdS black holes posses two extremal black holes in the range $\bb\in(1/6,\bex)$.

The heat capacity exhibits a negative and positive branch. It indicates a discontinuity in the heat capacity and electric permittivity, suggesting the occurrence a second-order phase transition that occurs at $T=T_{min}$ between the small and large branch. The denominator in both expressions determines a divergence at $S_0$ since it occurs when \eqref{s0} is satisfied\footnote{The phase transition order can be determined by using the Ehrenfest’s scheme by analyzing the divergence of heat capacity and electric permittivity, in the same way as it has been done in \cite{Banerjee:2011au} for RNAdS in higher dimensions.}. In contrast to the non-extremal RNAdS solution, for large potentials, namely $\Phi>\sqrt{2}$, the hairy black hole still presents two branches. Additionally, large black holes are locally stable, having both positive heat capacity and electric permittivity.
\bigbreak
\textbf{Case} {\boldmath$\gamma=\{0,-1\}$}: When the base manifold has a flat or hyperbolic topology, we observe the same qualitative behavior for the $T-S$ curves since all of them are monotonically increasing functions (see Fig. \ref{ts}). They have only one extremal black hole with entropy $S_1$, i.e. $T(S_1)=0$. The value of $S_1$ increases to infinity as we decrease $\bar{B}$ up to the flat case $\bb=1/6$ where there are no extremal configurations. The only difference can be noted in the strongly coupled limit by evaluating \eqref{tinf} for the respective topology. This determines a minimum possible value for the allowed entropies with positive temperatures, bounded from below by $S_1$ evaluated at the strongly coupled limit. Namely, $S_1>S^{\text{min}}_1$ with,
\be
S^{\text{min}}_1=-\frac{2\pi\gamma}{3}\ .
\ee
This means that the size of hyperbolic black holes is restricted by the topology of the base manifold. The hyperbolic and flat black holes are locally stable since their heat capacity and electric permittivity are positive for $S>S_1$.

\section{Phase transitions}\label{sex5}

The study of global stability determines which configuration is thermodynamically favored. This section's main idea is to compare the free energies of the hairy black hole, the RNAdS black hole and thermal AdS to determine which configuration is favored. Remind that these three configuration have the same asymptotics for $\bb>1/6$. We consider the grand canonical ensemble where the temperature $T$ and the electric potential $\Phi$ are fixed. In this ensemble, the Gibbs energy $\mathcal{G}$ for the hairy black hole is computed from \eqref{Ie} and \eqref{termosHBH} through the relation with the Euclidean action $\G=I_E/\beta=\M-TS-\Phi Q$, leading to the following expression,
\be\label{gibbs}
\G(T,\Phi)=\frac{\sqrt{S}}{24 \sqrt{2} \pi^{3 / 2}}\left(12 \pi \gamma+(1-6\bb)\frac{S(T,\Phi)}{\bb}+45 \times 5^{1 / 3} \pi^{2 / 3} \Phi^{4 / 3}\left(\frac{S(T,\Phi)}{\bb}\right)^{1 / 3} \right)\ .
\ee
Since it is not possible to get an explicit and analytical expression for $S(T,\Phi)$, we extract this relation numerically from \eqref{temp}. This is to fix $\Phi$ and $T$ in \eqref{temp} and solve for $S$ to get the numerical dependence. The free energy associated to the RNAdS can be analytically obtained as follows  \cite{Chamblin:1999tk}
\be
\G_0(T,\Phi)=\frac{1}{108}\left(4 \pi T\mp\sqrt{2} \sqrt{8 \pi^{2} T^{2}-6 \gamma+3 \Phi^{2}}\right)^{2}\left(2 \pi T\pm\sqrt{2} \sqrt{8 \pi^{2} T^{2}-6 \gamma+3 \Phi^{2}}\right)\ .
\ee
To compare the free energies, both black holes must be in the same grand canonical ensemble. This means that we must impose $T=T_0$ and $\Phi=\Phi_0$, where $T_0$ and $\Phi_0$ are the temperature and the electric potential of the RNAdS black hole, respectively. Thermal AdS corresponds to the vacuum configuration such that $G_{AdS}(T,\Phi)=0$ and, in consequence, we have three ``competing'' configurations in terms of thermodynamical stability. We proceed to study the global stability for each horizon topology.

\begin{figure}[H]
\centering
\includegraphics[width=.47\textwidth]{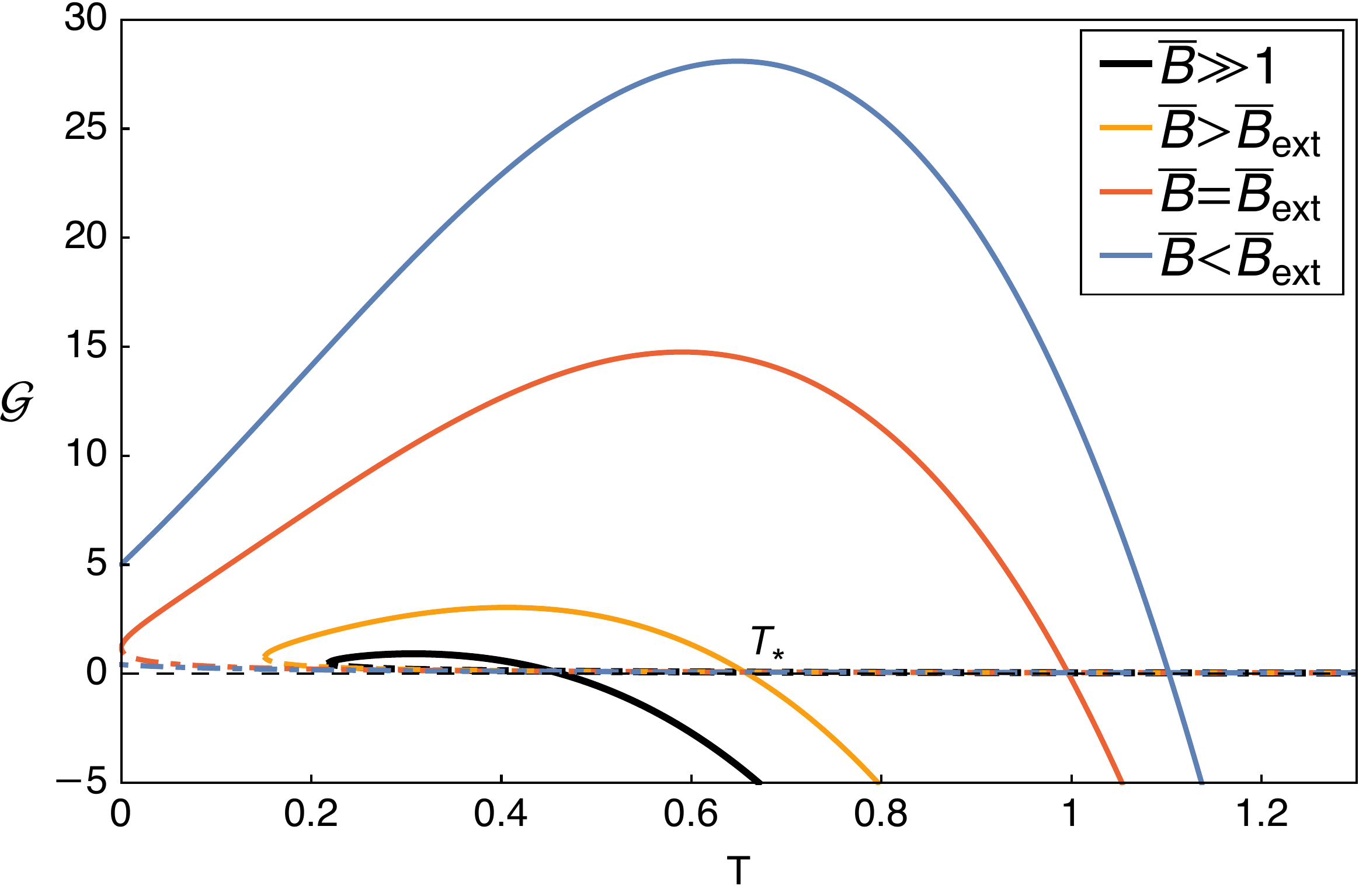}\hfill
\includegraphics[width=.47\textwidth]{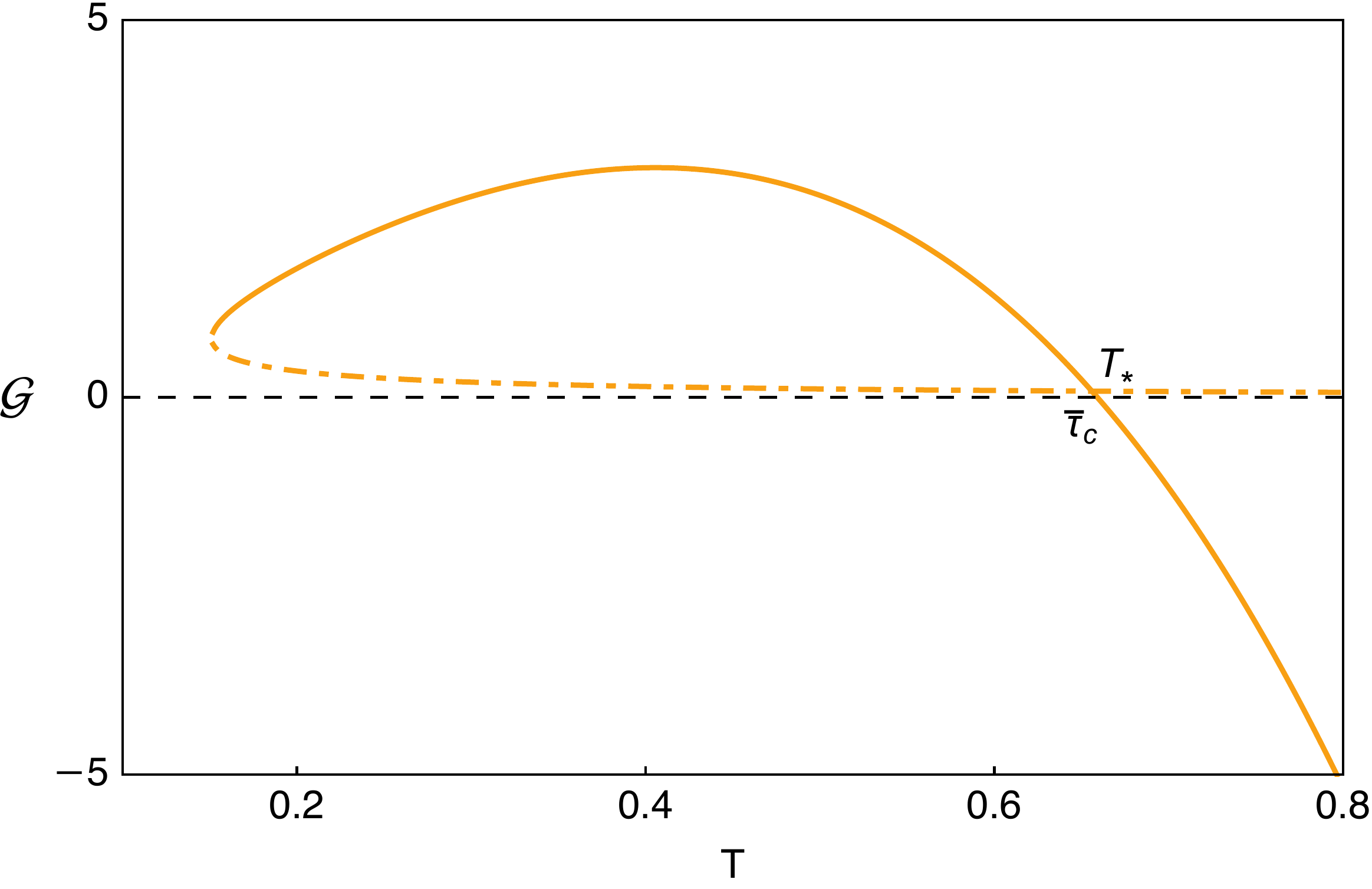}
\caption{Gibbs energy versus temperature for the hairy black hole with different values of the coupling parameter.The continuous line depicts the large branch and the dashed line the small branch of hairy black holes. Unlike the charged AdS RNAdS black hole in the grand canonical ensemble, there is no ``cusp" behavior. Instead, both branches meet at a second temperature reached at $\ts$, while a first-order phase transition can take place from thermal AdS to the large branch at $\tbc$. We have used the same values for the electric potential and $\bb$ than in Fig. \ref{ce}}.
\label{gt}
\end{figure}

\begin{figure}[H]
\centering
\includegraphics[width=.49\textwidth]{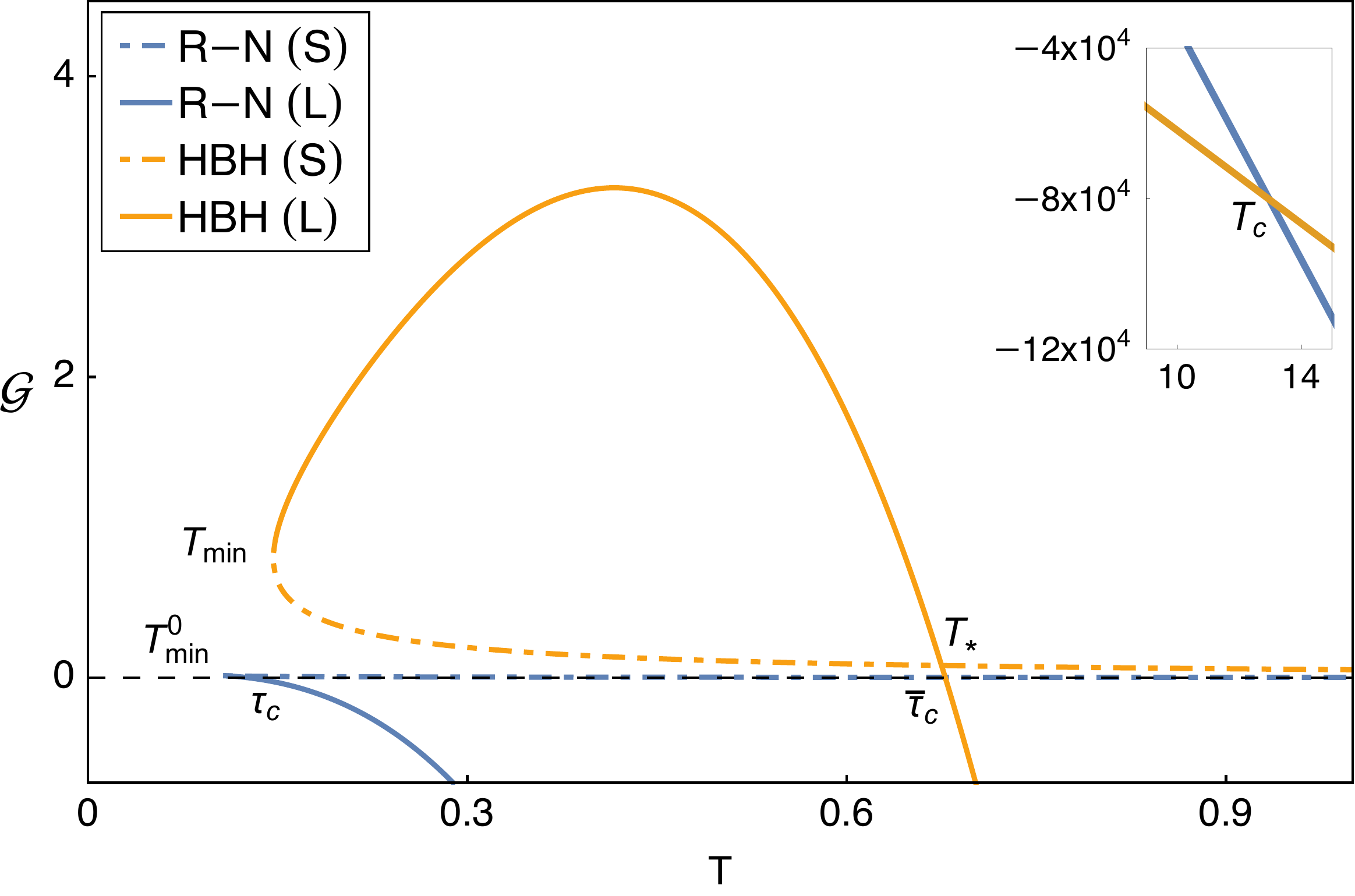}\hfill\includegraphics[width=.51\textwidth]{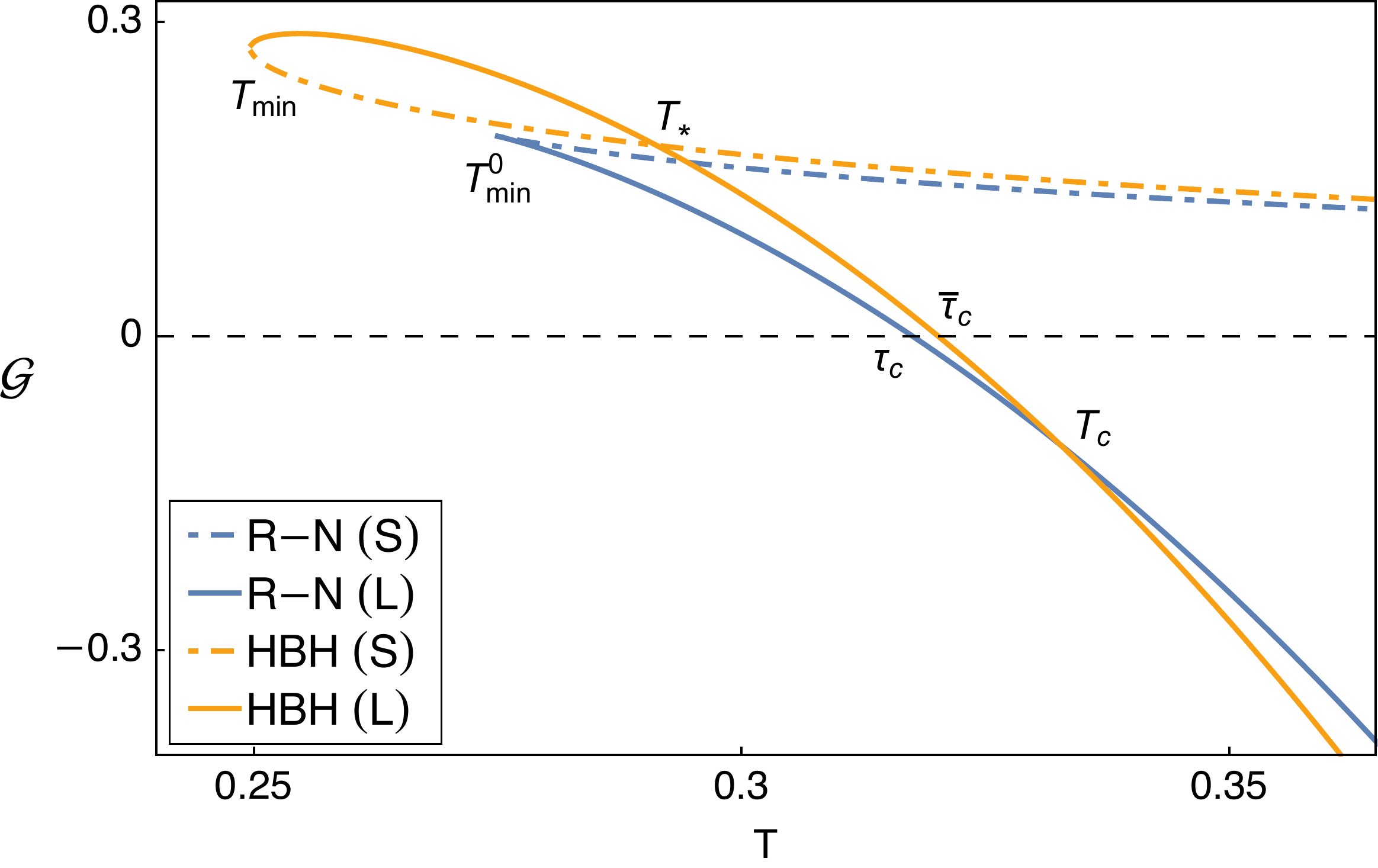}
\caption{Gibbs energy versus temperature for the hairy and Reissner-Nordstr\"om configurations: R-N (S) stands for the small RNAdS black holes, R-N (L) stands for the large RNAdS black holes, HBH (S) stands for the small hairy black holes and HBH (L) stands for the large hairy black holes. A first-order phase transition occurs at a temperature $T_c$ between large black holes from RNAdS to the hairy black holes. We have used $\Phi=1.3$ and $\bb=25$ and $\Phi=0.1$ and $\bb=1.5$ in the left and right panel, respectively.}
\label{case1}
\end{figure}

\begin{figure}[H]
\centering
\includegraphics[width=.5\textwidth]{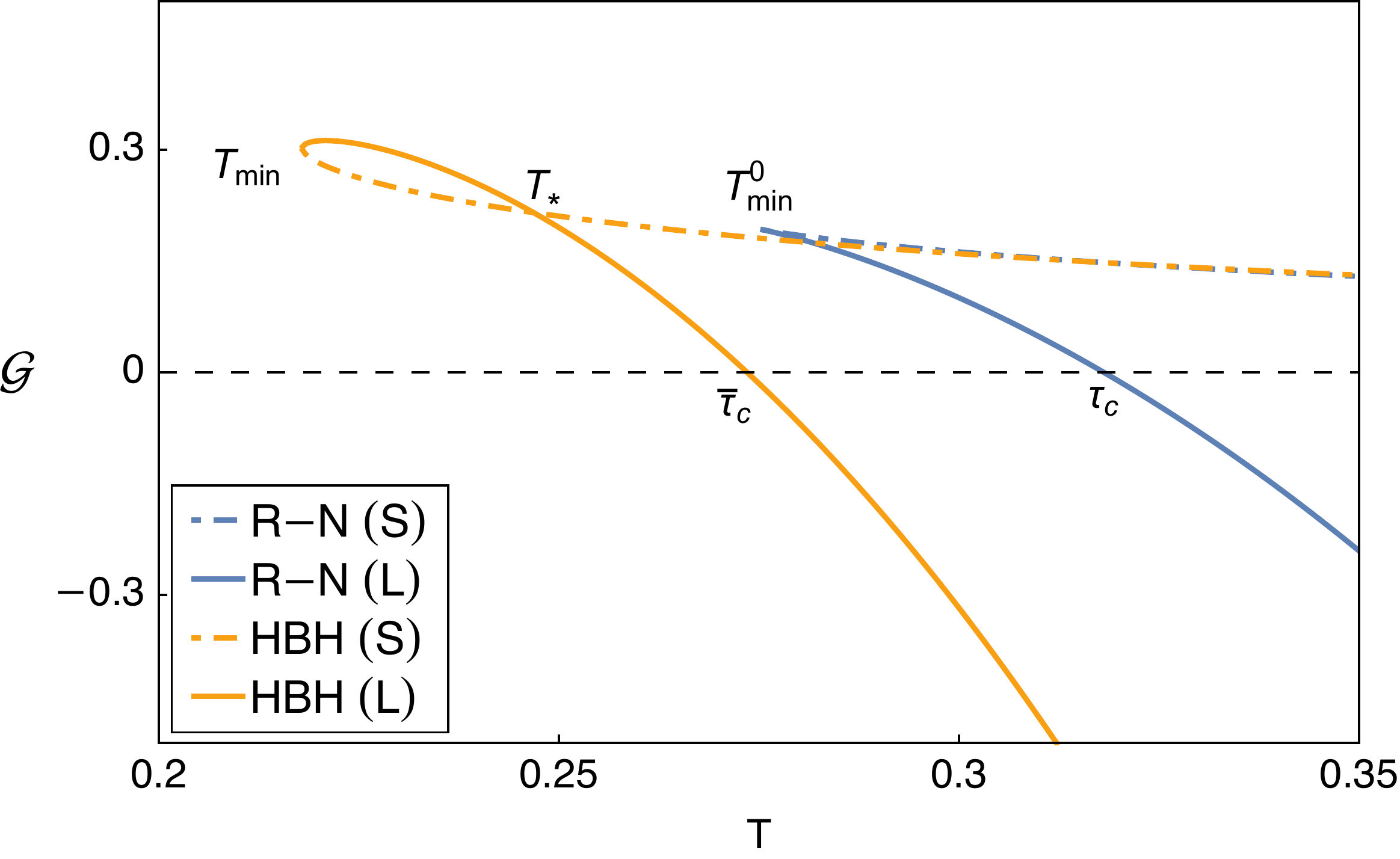}
\caption{Note that as we increase the temperature there is only one phase transition from thermal AdS to large hairy black holes at $\tbc$. We have used $\Phi=0.06$ and $\bb=0.5$.}
\label{case3}
\end{figure}

\begin{figure}[H]
\centering
\includegraphics[width=.5\textwidth]{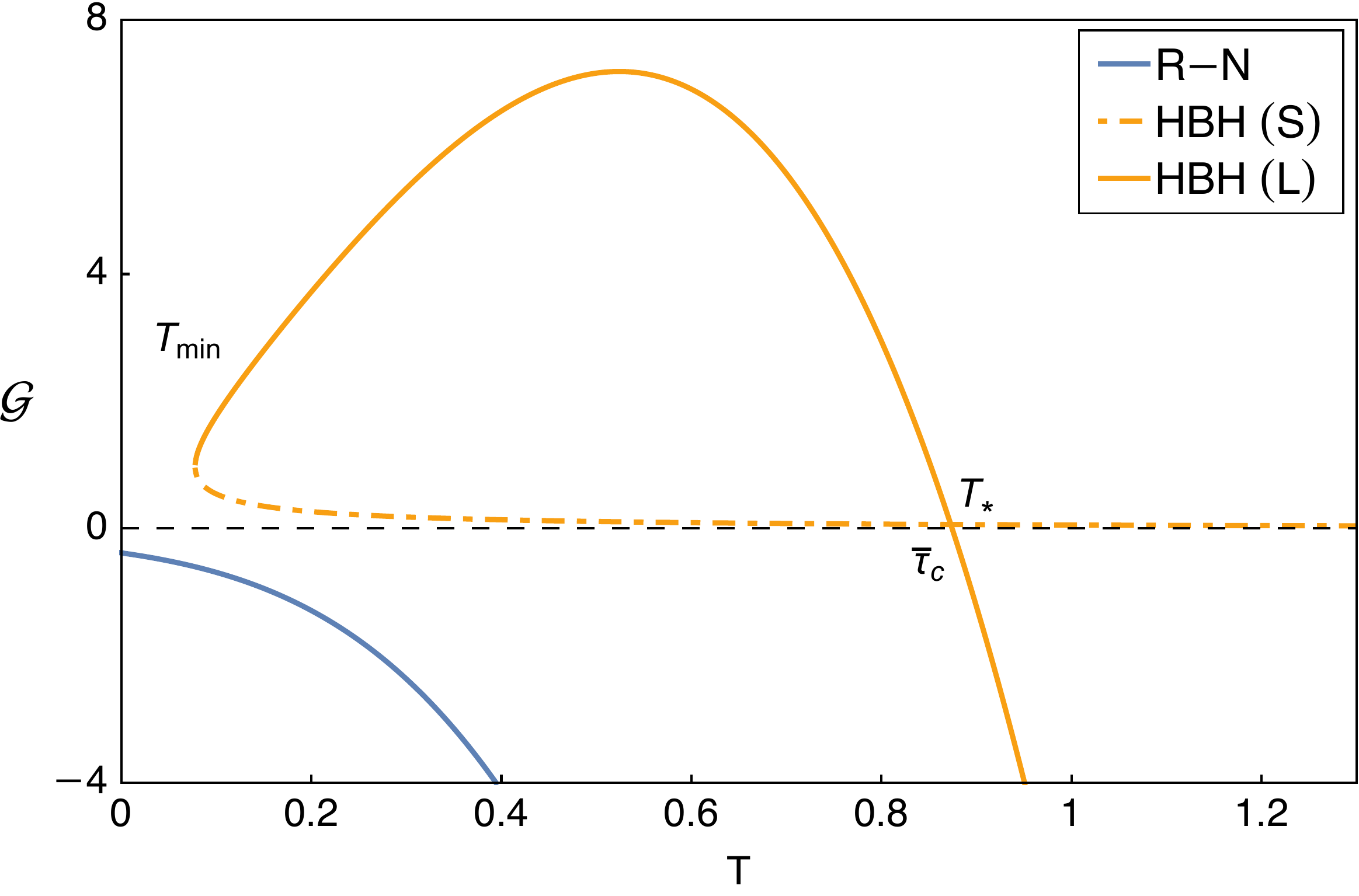}
\caption{For large potentials RNAdS black hole is the more stable configuration from $T=0$ to $T_c$, where there is a phase transition with the large branch of hairy black holes. We have used $\Phi=2$ and $\bb=40$.}
\label{case0}
\end{figure}

\bigbreak
\textbf{Case} {\boldmath$\gamma=1$}: Notice that in contrast with its GR counterpart, the Gibbs energy of the hairy black hole exhibits an intersection point at temperature $\ts$ between both branches and not a ``cusp" behavior. This means that the small branch is less probable compared to the large branch up to a temperature $\ts$ from which this behavior reverses. Then as we increase the temperature, a first-order phase transition can occur from thermal AdS to the large hairy black holes at a temperature $\tbc$ (see Fig. \ref{gt}).
For small electric potentials $\Phi<\sqrt{2}$, if $\tmv<\tm$, there are no available black hole configurations with temperatures $T<\tmv$ and thermal AdS is the favored configuration (see the left panel in Fig. \ref{case1}). For $\tmv\leq T<\tm$, RNAdS black holes can coexist in thermal equilibrium with thermal radiation. Still, they are less favored until it reaches a critical temperature $\tc$ where the first phase transition occurs, and the RNAdS black holes start dominating, being the most stable configuration. If we continue increasing the temperature, namely $\tm\leq T<T_c$, the hairy black hole configurations appear as possible states, but they are less favored than the RNAdS black hole, and large RNAdS black holes still dominate. It is not until a second phase transition at $T_c$, where the large hairy black holes dominate at higher temperatures over thermal AdS and RNAdS black holes.
A second situation is possible for small electric potentials when $\tm<\tmv<\ts$ (see the right panel in Fig. \ref{case1}). Here, the situation is the same except that hairy black hole configurations appear first, coexisting with pure AdS and then the RNAdS black holes join the ensemble. This happens as one increases the temperature from $\tm\leq T<\tmv$ and $\tmv\leq T<\tc$, respectively. For values of the coupling parameter $\bb \leq \bex$, two hairy black holes are possible configurations in the thermodynamic ensemble even at $T=0$, which correspond to the extremal ones.
 There is a third possible case for $\ts\leq\tmv$ where only one phase transition takes place at $\tbc$ from thermal AdS to the hairy black hole (see Fig. \ref{case3}). There are no phase transitions at large temperatures.
For large electric potentials $\sqrt{2}\leq\Phi$, the RNAdS configuration contains the extremal black hole (see Fig. \ref{case0}), dominating the partition function from $T=0$. As it is known, the RNAdS black hole has only one branch whose Gibbs energy is strictly negative, whereas the large branch of hairy black holes is the only branch having negative Gibbs energy for temperatures larger than $\tbc$. Interestingly enough, our numerical analysis determines that a first-order phase transition at higher temperatures $T_c$ compared to the case with small electric potential. As we increase even more the potential, $T_c$ gets larger and RNAdS dominates the partition function for a larger range of temperatures. The situations described are qualitatively the same when the hairy black hole contains the two extremal black holes in the range $\bb\leq\bex$.

From these cases, we can conclude that, as we increase the temperature, a first-order phase transition will take place such that the thermodynamic ensemble is dominated by a large hairy black hole either at $\tbc$ or $T_c$. To illustrate this, the phase structure for a fixed value of the coupling constant is summarized in the following diagrams:
\begin{figure}[H]
\centering
\includegraphics[width=.48\textwidth]{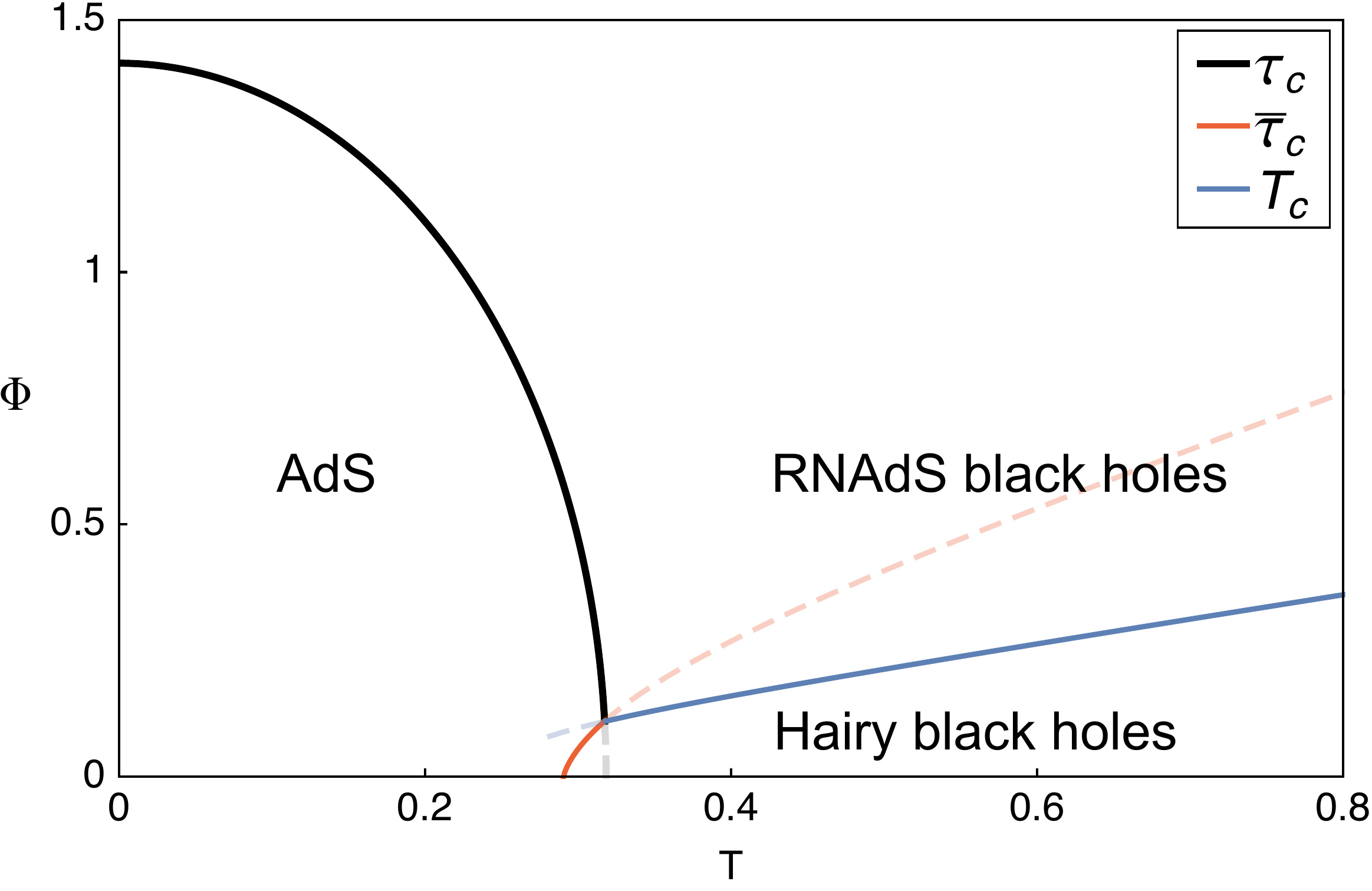}\hfill\includegraphics[width=.47\textwidth]{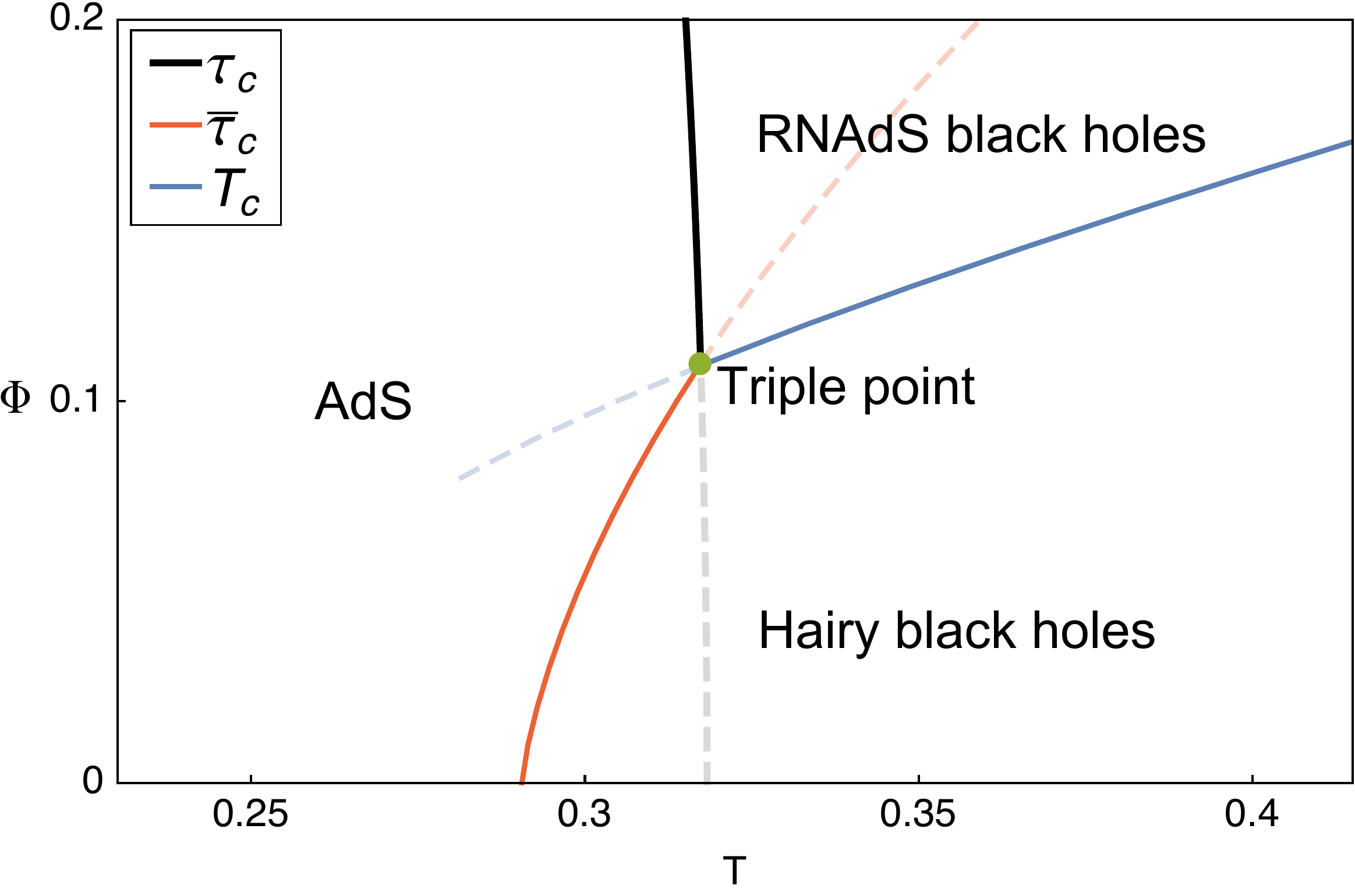}
\caption{Phase diagram $\Phi-T$. The phase diagram depicts the critical lines at which first-order phase transitions take place. The solid lines represents phase transitions between the most favored configurations. Each critical line are extended by dashed lines where first-order transitions occurs but at a Gibbs energy which is not the lowest one. The three lines converge at the triple point, as it can be seen in the magnified region at the right panel, which in this case is located at $T=0.317$ and $\Phi=0.110$ for $\bb=1$.}
\label{phd}
\end{figure}
The phase diagram depicts the critical lines at which first-order phase transitions take place. The solid lines represent phase transitions between the most favored configurations. Each critical line is extended by a dashed line where first-order transitions still occur but at higher Gibbs energies and therefore the favored configuration does not change. It is remarkable that the three possible states, thermal AdS, large RNAdS, and large hairy black holes, quite much resembles a solid-liquid-gas system, respectively, where the electric potential plays the role of pressure. One noticeable difference is that there is no critical point since the critical line for $T_c$ (liquid-gas critical line) extends to infinity, at least in the numerical domain from which we can infer that this is clearly the tendency. In close analogy to a solid-liquid-gas thermodynamical system, there is a triple point where three phases coexist being equally probable. It is particularly interesting the evolution of each region in the phase diagrams as a function of the coupling parameter as one goes from the flat limit $\bb=1/6$ to the strong coupling regime. In Fig. \ref{phd}, as we increase the value of $\bb$, the critical point rapidly descends over the $\tc$ critical line in a clockwise fashion, while the hairy black hole region that enters into the AdS shrinks. This abrupt behavior can be seen if we collect the critical temperatures for many values of $\bb$, mainly around the asymptotically flat limit ($\bb\gtrsim 1/6$). We obtained the following result,
\begin{figure}[H]
\centering
\includegraphics[width=.473\textwidth]{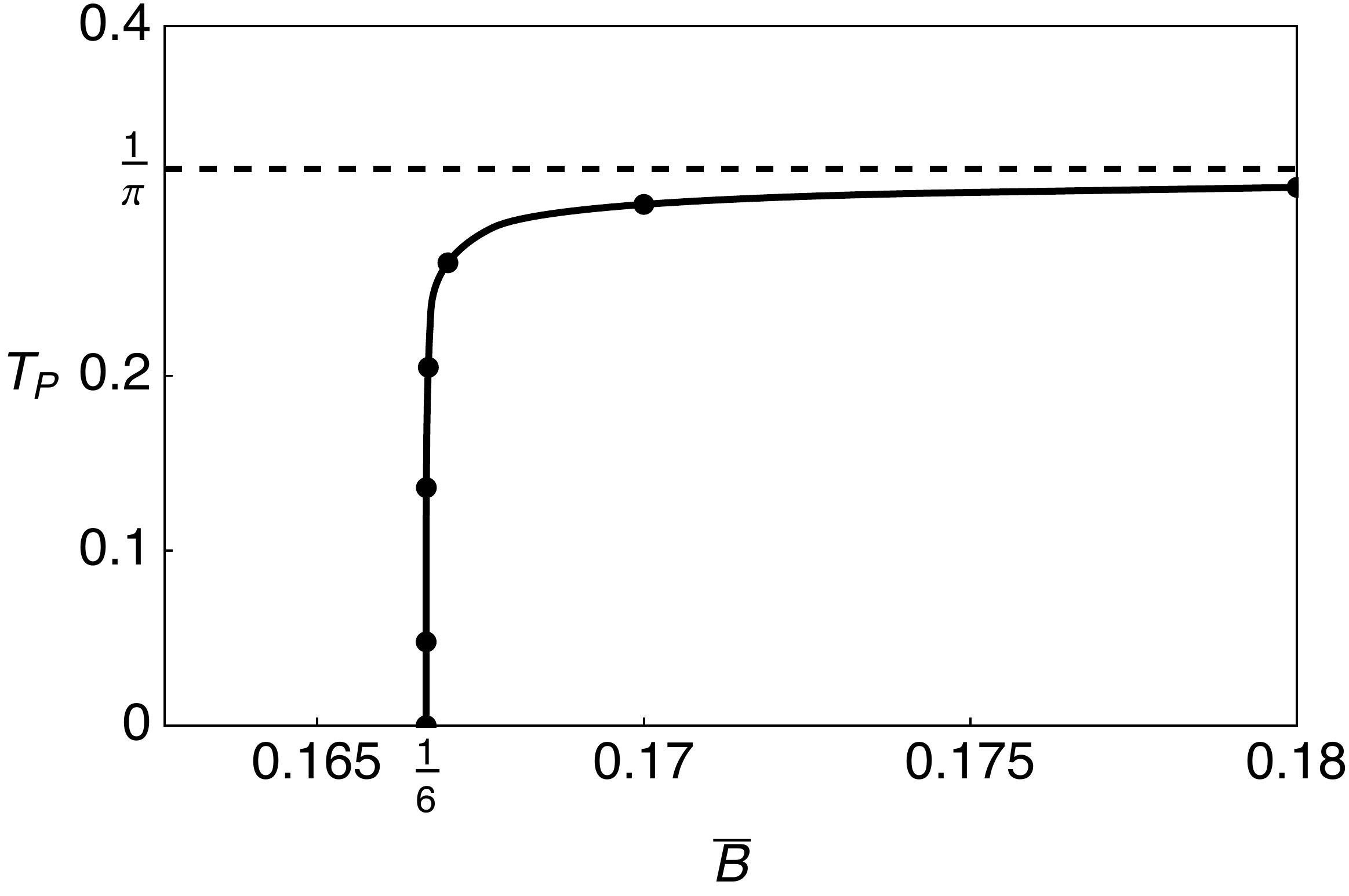}\hfill\includegraphics[width=.483\textwidth]{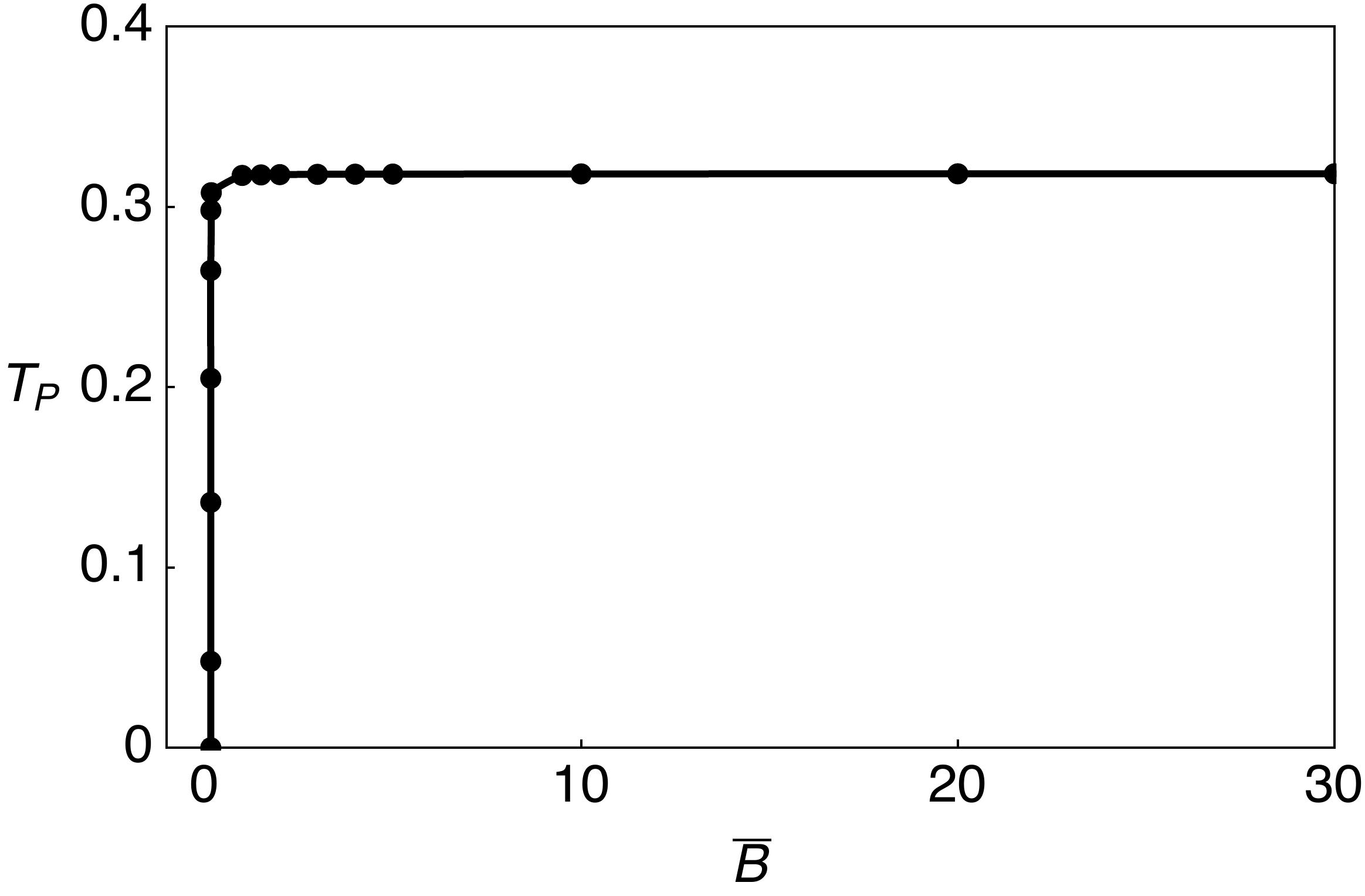}
\caption{Temperatures of triple points as a function of the coupling parameter. This curve presents a steep slope close to the asymptotically flat limit, and rapidly tends to the Schwarzschild critical temperature. Recall that the asymptotically flat limit is the value of the coupling parameter at which the effective cosmological constant tends to zero. In consequence, the hairy black holes abruptly reduce their zone of dominance in the phase diagram, as soon as the coupling constant departs from the flat limit.}
\label{tp}
\end{figure}
For values of $\bb\gtrsim1/6$ the $\tbc$ and $T_c$ curves get closer, straighter, and almost vertical, meaning that in the weakly coupled regime, the phase diagram is mainly dominated by the large hairy black hole phase. A particularly simple and analytical expression for this limit can be obtained using series expansions in \eqref{temp} and \eqref{gibbs}, getting,
\be\label{tbceq}
\tbc(\bb\gtrsim1/6)\sim\frac{5}{\pi}(6\bb-1)^{1/4}\Phi\ ,
\ee
which exhibits the described behavior.
On the other hand, as we increase the scalar coupling, the AdS region and large RNAdS phases rapidly dominate the phase diagram, and curiously enough, the limit of $\bb\rightarrow\infty$ tends to a critical temperature which is exactly the critical temperature of Schwarzschild black holes. This can be seen by the analytical expression of $\tbc$ at $\Phi=0$, which is,
\be
\tbc(\Phi=0)=\frac{1}{\pi}\sqrt{1-\frac{1}{6\bb}}\ .
\ee

Note that phase transitions at $T_c$ occur only if the electric potential reaches a minimum value, which is why this critical line enters to the AdS region, but it has an ending point. For lower values than that minimum, we are in the situation depicted in Fig. \ref{case3}.

\begin{figure}[H]
\centering
\includegraphics[width=.48\textwidth]{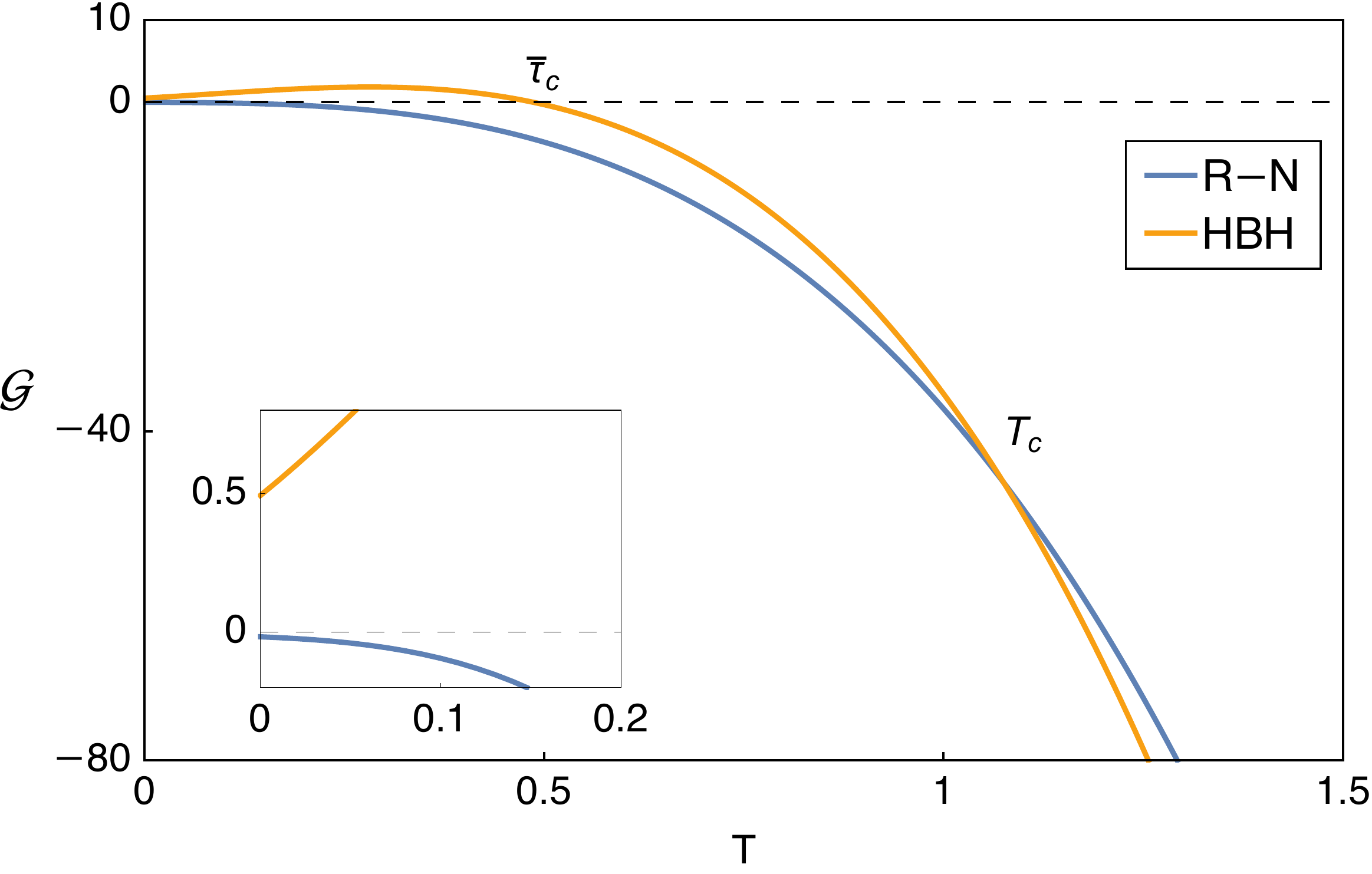}\hfill\includegraphics[width=.47\textwidth]{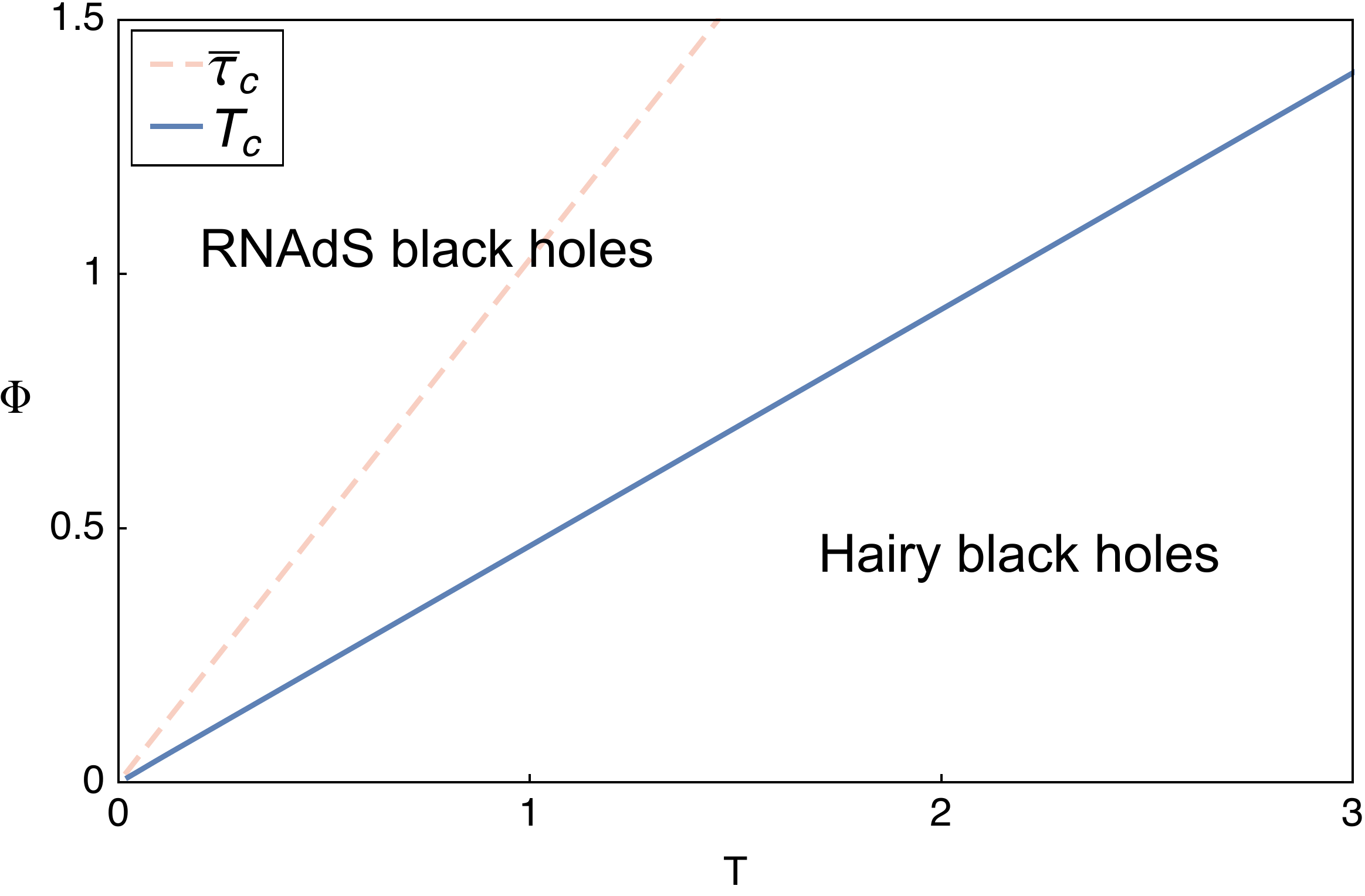}
\caption{Left panel: Gibbs energy versus temperature for the hairy black hole and Reissner-Nordstr\"om configurations. We have considered $\Phi=0.5$ for $\bb=1$. Right panel: Phase diagram of the two possible favored configurations. The dashed line represents phase transition at higher Gibbs energy. Here we have used $\bb=1$.}
\label{plotg0}
\end{figure}

\bigbreak
\textbf{Case} {\boldmath$\gamma=0$}: The flat hairy black hole, as well as the RNAdS black hole, have only one branch containing only one extremal black hole. The former possesses positive Gibbs energy and the latter negative Gibbs energy in their extremal configurations. The ensemble is dominated from $T=0$ up to $T_c$ by the RNAdS black hole since its Gibbs energy conserves a negative value, and only a first-order phase transition occurs at $T_c$ from which the hairy black hole starts to dominates. This is independent of the value on the space of parameters $\bb-\Phi$ as it can be seen in the phase diagram (see the right panel in Fig. \ref{plotg0}).
Note in the left panel of Fig. \ref{plotg0} that there are no phase transitions between thermal AdS and RNAdS. There are only two phase transitions, the first one at $\tbc$ which is at higher Gibbs energy, and in consequence, it does not change the favored configuration while a second one occurs at higher temperature $T_c$ that switches the dominance of the thermodynamics in favor of the hairy black hole. Like the spherical case, as we approach $\bb=1/6$, the zone dominated by the hairy black hole in the phase diagram can be drastically increased, while the critical line $T_c$ gets closer to the critical line $\tbc$. The latter behaves in the same way as equation \eqref{tbceq} describes.

\bigbreak
\textbf{Case} {\boldmath$\gamma=-1$}: The hyperbolic black hole has qualitatively similar behavior to the flat case, although there is a particular property that makes a slight difference. Like the flat case, thermal AdS does not have any dominance in the phase diagram. However, the thermodynamic ensemble can be dominated for small electric potentials even from zero temperature by the extremal black hole.
\begin{figure}[H]
\centering
\includegraphics[width=.33\textwidth]{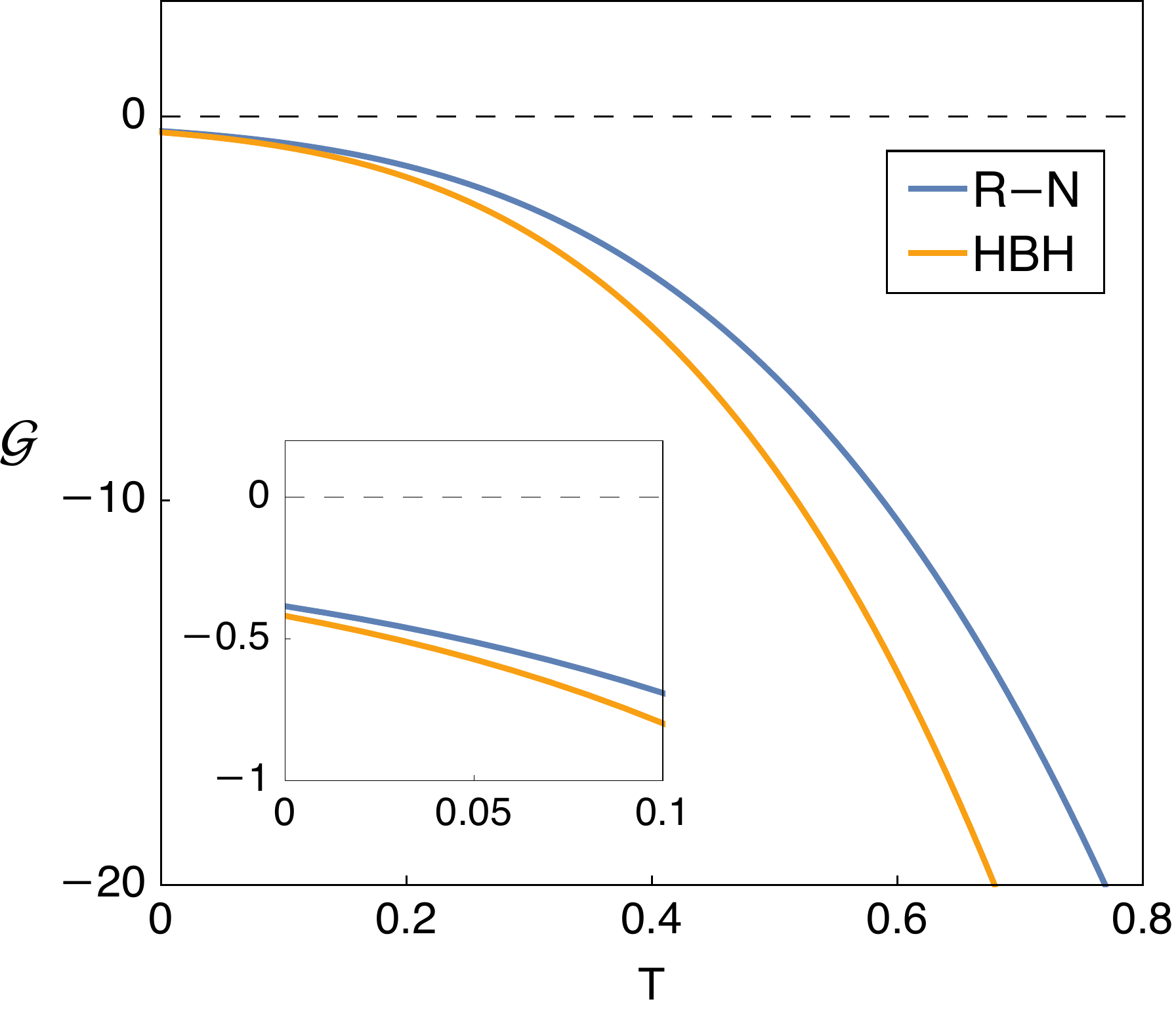}\hfill\includegraphics[width=.33\textwidth]{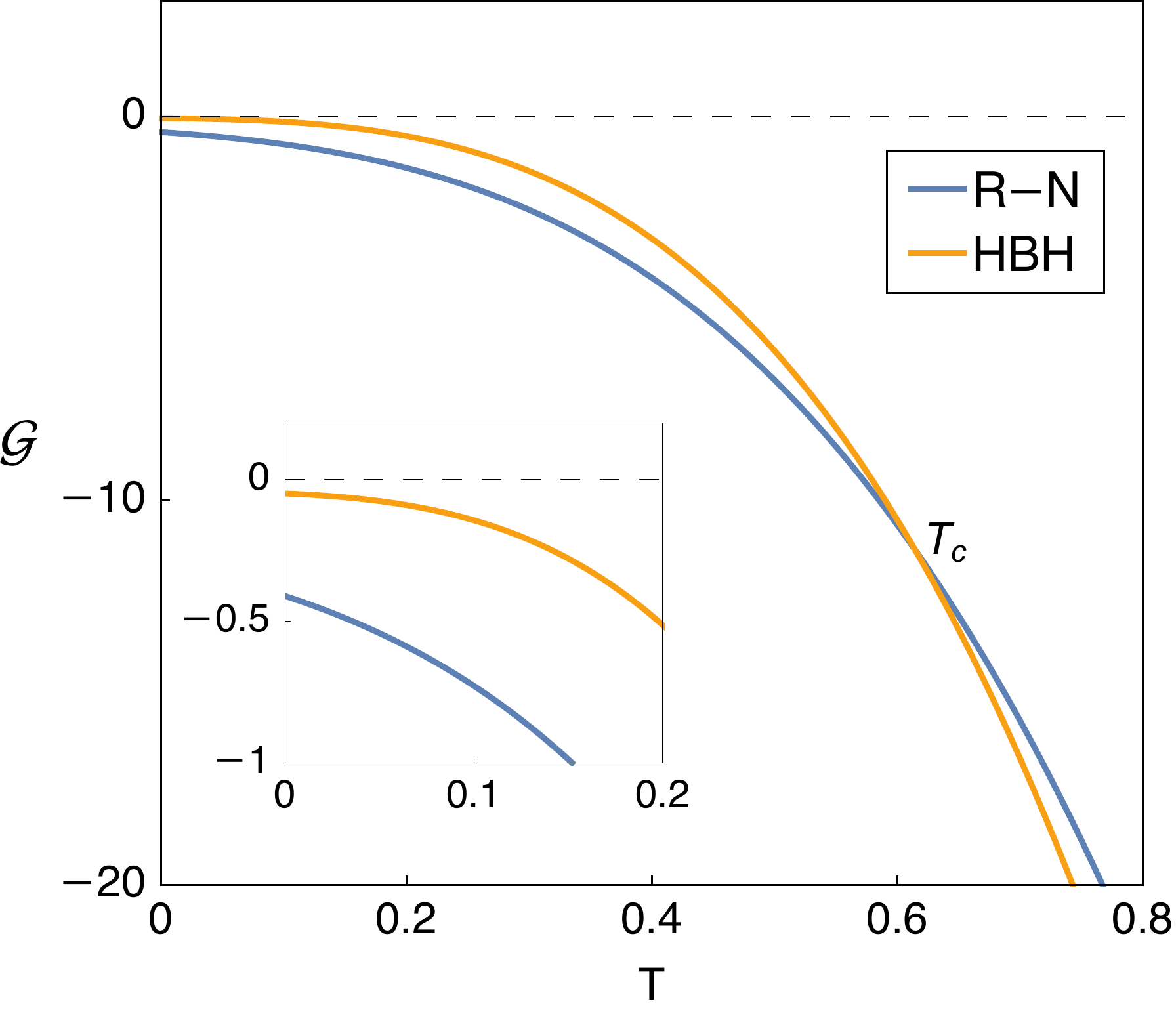}\hfill\includegraphics[width=.33\textwidth]{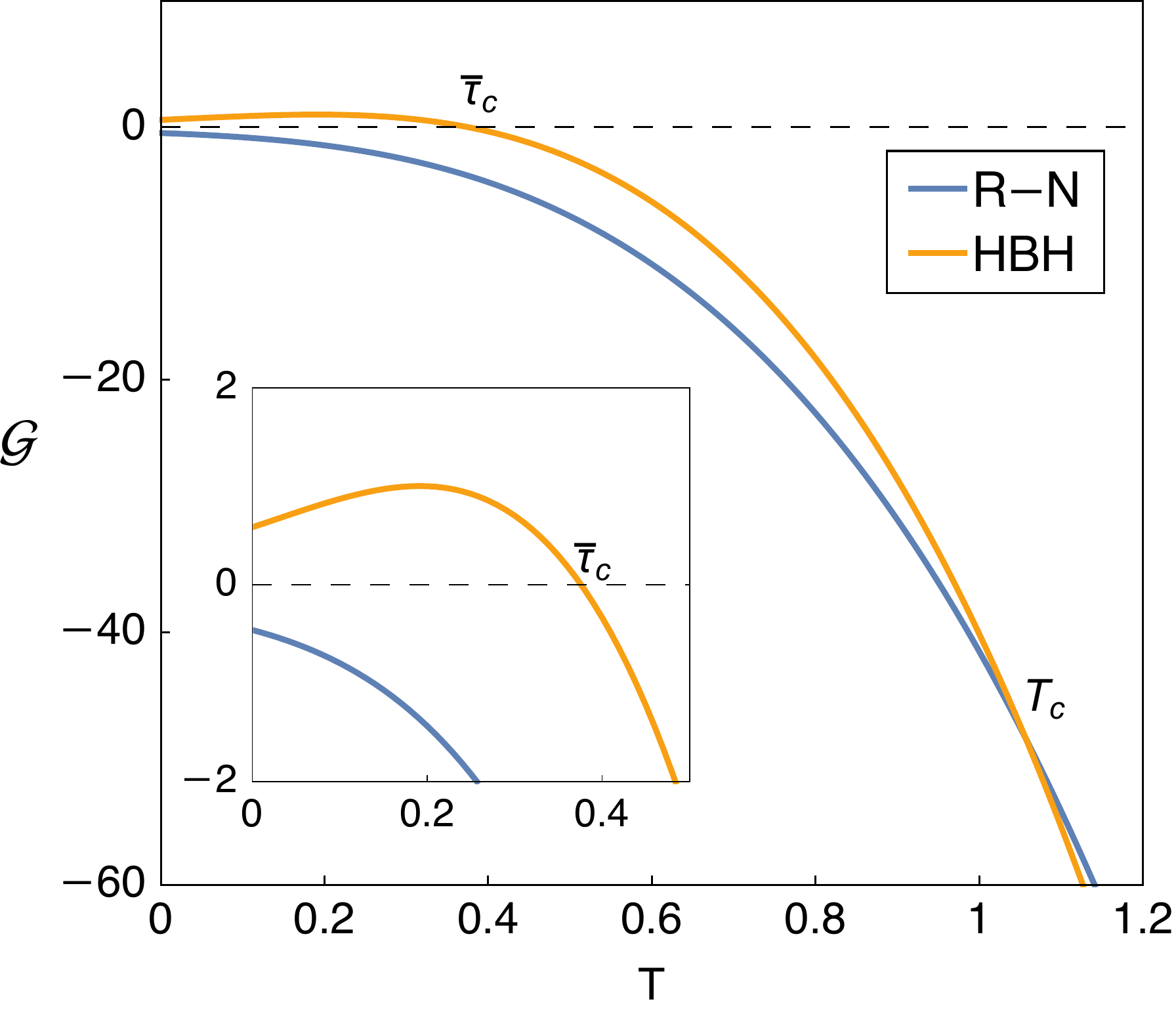}
\caption{Gibbs energy versus temperature for the hairy black hole and Reissner-Nordstr\"om configurations. From left to right, we have considered $\Phi=0.01$, $\Phi=0.3$ and $\Phi=0.5$ for $\bb=1$.}
\label{plotgm}
\end{figure}
In Fig. \ref{plotgm}, and from left to right, we describe three possible behaviors in the phase structure as we increase the electric potential. The left plot shows a thermodynamic system under a small electric potential, where the hairy black hole has the global minimum of Gibbs energy for all temperatures. As shown in the plot at the middle, once the electric potential reaches a particular value, the system allows RNAdS as a favored configuration from extremal black holes up to a phase transition at $T_c$ where the hairy black hole recovers its dominance on the ensemble. Finally, as the right plot describes, if we continue increasing the electric potential, the extremal hairy configuration can get a positive Gibbs energy and allows phase transitions to thermal AdS at $\tbc$. Still, it is not enough to modify the favored configurations for the previous situation.
\begin{figure}[H]
\centering
\includegraphics[width=.5\textwidth]{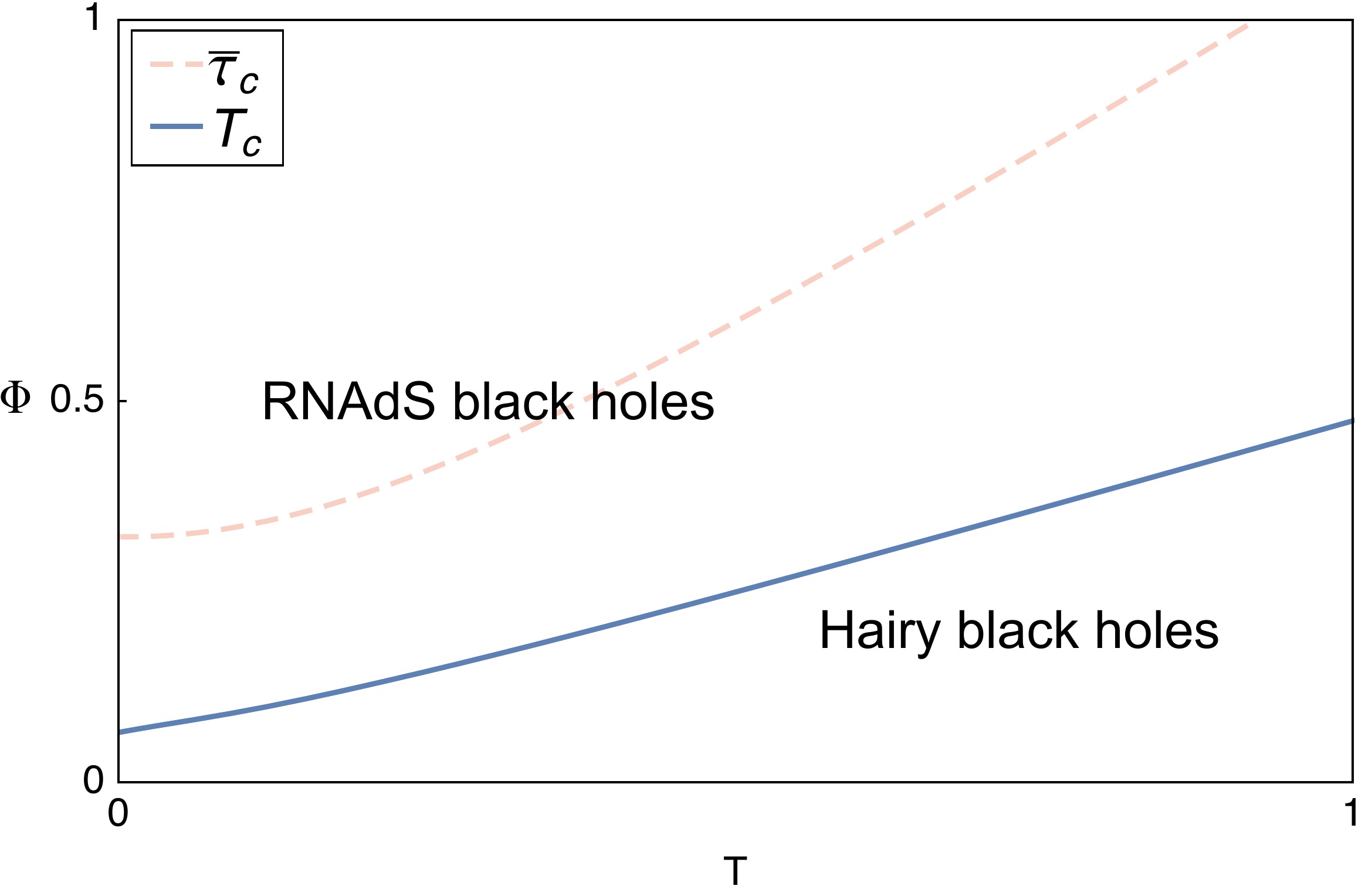}
\caption{Phase diagram of the two possible favored configurations. The dashed line represents phase transition at higher Gibbs energy. Here we have used $\bb=1$.}
\label{phm}
\end{figure}
Like the flat black holes, as we decrease the coupling parameter, the region of hairy black holes in the phase diagram increases.

\section{Conclusions and discussion}\label{sex6}

We have derived black hole solutions and studied their thermodynamic behavior in bi-metric scalar-tensor theories with disformal symmetry. The main difficulty in these theories is the existence of no-hair obstructions, a direct consequence of the scalar field present in this kind of theory. Inspired by the no-hair theorem for galileons and exploiting the shift invariance of the theory, such obstruction can be evaded by requiring a regular conserved current associated with the shift symmetry, a fact that manifests in the scalar field equation. As a consequence, we obtained an exact electrically charged black hole solution in the presence of a real scalar field that is regular at and outside the horizon. A curvature singularity at the origin is hidden by an event horizon which is a surface of spherical, flat, or hyperbolic topology.

The coupling parameter of the scalar field introduces an effective cosmological constant which determines the black hole asymptotics allowing the black hole solutions to exhibit AdS, flat, or dS asymptotic behavior. Because of the presence of the electric charge, the theory also admits the Reissner-N\"{o}rdstrom solution with a trivial scalar field, a suitable configuration to be considered in the thermodynamic analysis. This is carried out by employing the Euclidean approach, where this solution can be regarded as a third configuration, along with thermal AdS and the hairy black hole as possible states of the grand canonical ensemble by keeping fixed the temperature and the electric potential.

 Focused on the AdS asymptotics, we have found rich thermodynamics whose behavior is highly dependent on the coupling strength of the scalar field and the topology of the horizon. In contrast to its GR counterpart, the spherical configuration presents branches with large and small black holes even for large electric potential. All the topologies admit large black holes which are locally stable, meaning that they are stable under thermal and electric fluctuations. The global stability is analyzed by comparing Gibbs energies between the three possible configurations. To achieve this, we have adopted the grand canonical ensemble, i.e., an ensemble at fixed electric potential and temperature, and we investigated the consequences of the coupling parameter on the phase structure. Whether a configuration is thermodynamically preferred or not must be determined considering its Gibbs energy relative to other configurations at the same electric potential and temperature.

 In the spherical case, the Gibbs energy of the hairy black hole exhibits an intersection point at a specific temperature between both branches and not a ``cusp" behavior. Unlike the RNAdS black hole, small black holes can be more probable than large black holes undergoing a first-order phase transition at the intersection point where this behavior reverses. If we continue increasing temperature, a transition from thermal AdS to large black holes takes place. However, the competition by the minimum Gibbs energy is modified once RNAdS is taking into account. Strikingly, from a thermodynamic perspective, the phase structure resembles a solid-liquid-gas system (see Fig. \ref{phd}), where the electric potential plays the role of pressure. In close analogy, there is a triple point where the three phases coexist, being equally probable.

In contrast to the spherical black hole, flat black holes and flat RNAdS black holes have only one branch. Since the latter have always negative Gibbs energy while the former contains extremal black holes at positive Gibbs energy, this one undergoes two phase transitions; the first one from thermal AdS to the hairy black hole which is at higher Gibbs anergy, and in consequence, it does not change the favored configuration, and a second one at a higher temperature that finally switches the dominance of the thermodynamics in favor of the hairy black hole. The phase structure is significantly simple compared to the spherical case since there are only two preferred phases. The phase diagram in the hyperbolic case is similar to the flat case but with one particularly curious distinction. If the electric potential does not exceed a certain threshold potential, the thermodynamic system is dominated by hairy black holes even at zero temperature. Above that threshold, the phase diagram starts to share the phase space with RNAdS black hole.

We found that strongly coupled systems abruptly increase the predominance of the undressed black holes in the phase space (see Fig. \ref{tp}), while the weak regime increases the predominance of the hairy configuration. We have only studied the asymptotically AdS black holes; it would be interesting to explore how this behavior and the phase structure can be modified when the coupling confers flat or dS asymptotics. On the other hand, since the solution contains horizons with different topologies, the flat geometry of the horizon opens the possibility to study holographic applications based on the gauge/gravity duality. Including a mechanism of momentum dissipation DC conductivities and Hall angle of the holographic theory dual to the hairy black hole can be studied to determine the effect of the coupling parameter in this context.

\section*{ACKNOWLEDGMENTS}

C. E. acknowledges financial support given by PAI Grant No. 77190046. C.E also thanks to National Technical University of Athens for warm hospitality and support during the early stage of this work.


\end{document}